\documentclass[preprint,3p,12pt]{elsarticle}
\usepackage{amsmath}
\usepackage{stmaryrd}
\usepackage{bbding}
\usepackage{dcolumn}
\usepackage{graphicx}
\usepackage{amsfonts}
\usepackage{amssymb}
\usepackage{psfrag}
\usepackage{wrapfig}
\usepackage{subfigure}
\usepackage{makeidx}
\usepackage{bm}
\usepackage{epsf}
\usepackage{epsfig}
\usepackage{setspace}
\usepackage{color}

\begin{document}
\title{High-order ALE gas-kinetic scheme with unstructured WENO reconstruction}
\author[BNU]{Liang Pan\corref{cor}}
\ead{panliang@bnu.edu.cn}
\author[HKUST1]{Fengxiang Zhao}
\ead{fzhaoac@connect.ust.hk}
\author[HKUST2,HKUST1,HKUST3]{Kun Xu}
\ead{makxu@ust.hk}
\address[BNU]{School of Mathematical Sciences, Beijing Normal University, Beijing, China}
\address[HKUST1]{Department of Mechanical and Aerospace Engineering, Hong Kong University of Science and Technology, Kowloon, Hong Kong}
\address[HKUST2]{Department of mathematics, Hong Kong University of Science and Technology, Kowloon, Hong Kong}
\address[HKUST3]{Shenzhen Research Institute, Hong Kong University of Science and Technology, Shenzhen,
China}
\cortext[cor]{Corresponding author}

\begin{abstract}
In this paper, a high-order multi-dimensional gas-kinetic scheme is
presented for both inviscid and viscous flows in arbitrary
Lagrangian-Eulerian (ALE) formulation. Compared with the traditional
ALE method, the flow variables are updated in the finite volume
framework, and the rezoning and remapping steps are not required.
The two-stage fourth-order method is used for the temporal
discretization, and the second-order gas-kinetic solver is applied
for the flux evaluation. In the two-stage method, the spatial
reconstruction is performed at both initial and intermediate stage,
and the computational mesh at the corresponding stage is given by
the specified mesh velocity. In the moving mesh procedure, the mesh
may distort severely and the mesh quality is reduced. To achieve the
accuracy and improve the robustness, the newly developed WENO method
\cite{un-WENO3} on quadrilateral unstructured meshes is adopted at
each stage. The Gaussian quadrature is used for flux calculation.
For each Gaussian point, the reconstruction performed in the local
moving coordinate, where the variation of mesh velocity is taken
into account. Therefore, the accuracy and geometric conservation law
can be well preserved by the current scheme even with the largely
deforming mesh. Numerical examples are presented to validate the
performance of current scheme, where the mesh adaptation method and
cell centered Lagrangian method are used to provide mesh velocity.
\end{abstract}


\maketitle
\section{Introduction}
In the computational fluid dynamics, there exist two coordinate
systems to describe flow motions, where the Eulerian system
describes flow motions with time-independent mesh and the mesh moves
with the fluid velocity in the Lagrangian system. Considerable
progress has been made over the past decades for the numerical
simulations based on these two descriptions
\cite{Lagrangian,mesh-vel-1,Riemann-appro}. The Eulerian system is
relatively simple, but it smears contact discontinuities and slip
lines badly. For the computation with fixed boundaries, the Eulerian
system could work effectively. However, in the problems with moving
boundaries and bodies, it becomes difficult for the Eulerian method.
On the other hand, the Lagrangian system could resolve contact
discontinuities sharply, but computation can easily break down due
to mesh deformation.  In order to avoid mesh distortion and tangling
in the Lagrangian method, a widely used
arbitrary-Lagrangian-Eulerian (ALE) technique was developed
\cite{ALE1}, which contains three main steps. In the Lagrangian
stage, the solution and the computational mesh are updated
simultaneously. To release the error due to mesh deformation, the
computational mesh is adjusted to the optimal position in the
rezoning stage. In the remapping stage, the Lagrangian solution is
transferred into the rezoned mesh. Further developments and great
achievement have been made for the  ALE method
\cite{ALE2,ALE3,ALE4,ALE5}.

In the past decades, the gas-kinetic scheme (GKS) based on the
Bhatnagar-Gross-Krook (BGK) model \cite{BGK-1,BGK-2} has been
developed systematically for computations from low speed flow to
supersonic one \cite{GKS-Xu1,GKS-Xu2}. Different from the numerical
methods based on the Riemann flux \cite{Riemann-appro}, GKS presents
a gas evolution process from kinetic scale to hydrodynamic scale,
where both inviscid and viscous fluxes are recovered from a
time-dependent and multi-dimensional gas distribution function at a
cell interface. Recently, based on the time-dependent flux function
of the generalized Riemann problem (GRP) \cite{GRP1,GRP2} and
gas-kinetic scheme, a two-stage fourth-order method was developed
for Lax-Wendroff type flow solvers, particularly applied for the
hyperbolic conservation laws \cite{GRP-high-1, GRP-high-2,
GKS-high-1, GKS-high-2, GKS-high-3}. Under the multi-stage
multi-derivative framework, a reliable two-stage fourth-order GKS
has been developed, and even higher-order of accuracy can be
achieved. More importantly, this scheme is as robust as the
second-order scheme and works perfectly from subsonic to hypersonic
flows.  Based on the unified coordinate transformation \cite{Hui},
the second-order gas-kinetic scheme was developed under the
moving-mesh framework \cite{Jin1,Jin2}. With the integral form of
the fluid dynamic equations, a second-order gas-kinetic scheme with
arbitrary mesh velocity was developed \cite{GKS-ALE}, in which the
rezoning and remapping step in the traditional ALE method is
avoided. It provides a general framework for the moving-mesh method,
which can be considered as a remapping-free ALE-type method. The
piecewise constant mesh velocity at each cell interface is
considered in the ALE-type gas-kinetic scheme, which is reasonable
for a second-order scheme. However, for the higher-order schemes,
the variation of the mesh velocity along a cell interface needs to
be taken into account. Otherwise, the geometric conservation law may
be violated during the mesh rotating and deformation. Recently, a
one-stage DG-ALE gas-kinetic method was developed, especially for
the oscillating airfoil calculations \cite{DG-GKS}. The variation of
mesh velocity is considered in the time dependent flux calculation
along a cell interface, and the geometric conservation law can be
satisfied accurately. However, the one-stage gas evolution model and
DG framework become very complicated, and the efficiency of the
DG-ALE-GKS becomes low due to the severely constrained time step
from the DG formulation.

In this paper, a robust WENO scheme \cite{un-WENO3}, which was
developed recently on unstructured quadrilateral meshes, is extended
into the gas-kinetic framework. The optimization approach for linear
weights and the non-linear weights with new smooth indicator
improves the robustness of traditional WENO scheme. With the
arbitrary Lagrangian-Eulerian formulation, a high-order moving-mesh
gas-kinetic scheme is developed for the inviscid and viscous flows.
It extends the gas-kinetic scheme with unstructured WENO
reconstruction from the static domain to the variable one. The
two-stage fourth-order method is used for the temporal
discretization, in which the computational mesh at the corresponding
stage is given by the specified mesh velocity, and the WENO
reconstruction is performed at the initial and intermediate stage
respectively. The mesh velocity inside each cell interface are
considered and the WENO reconstruction is performed at each Gaussian
quadrature point in the local moving coordinate. Thus, the accuracy
and geometric conservation law can be well preserved by the current
scheme even with the largely deforming mesh. Numerical examples are
presented to validate the performance of current scheme, where the
mesh adaptation method and cell centered Lagrangian method are used
to provide the mesh velocity.

This paper is organized as follows. In Section 2, the gas-kinetic
scheme on moving-mesh is introduced. The extension of the two-stage
temporal discretization to moving mesh system is presented in
Section 3. In Section 4, we give  the WENO reconstruction on
unstructured quadrilateral meshes. Section 5 includes numerical
examples to validate the current algorithm. The last section is the
conclusion.

\section{Gas-kinetic scheme on moving-meshes}
The two-dimensional BGK equation can be written as \cite{BGK-1,BGK-2},
\begin{equation}\label{bgk}
f_t+uf_x+vf_y=\frac{g-f}{\tau},
\end{equation}
where $f$ is the gas distribution function, $g$ is the corresponding
equilibrium state, and $\tau$ is the collision time. The collision
term satisfies the compatibility condition
\begin{equation}\label{compatibility}
\int \frac{g-f}{\tau}\psi \text{d}\Xi=0,
\end{equation}
where $\psi=(1,u,v,\displaystyle \frac{1}{2}(u^2+v^2+\xi^2))$,
$\text{d}\Xi=\text{d}u\text{d}v\text{d}\xi^1...\text{d}\xi^{K}$, $K$
is the number of internal degrees of freedom, i.e.
$K=(4-2\gamma)/(\gamma-1)$ for two-dimensional flows, and $\gamma$
is the specific heat ratio. In the continuum regime, the gas
distribution function can be expanded as
\begin{align*}
f=g-\tau D_{\boldsymbol{u}}g+\tau D_{\boldsymbol{u}}(\tau
D_{\boldsymbol{u}})g-\tau D_{\boldsymbol{u}}[\tau
D_{\boldsymbol{u}}(\tau D_{\boldsymbol{u}})g]+...,
\end{align*}
where $D_{\boldsymbol{u}}={\partial}/{\partial
t}+\boldsymbol{u}\cdot \nabla$. With the zeroth order truncation,
i.e. $f=g$, the Euler equations can be obtained.  For the
Navier-Stokes equations, the first order truncation is used
\begin{align*}
f=g-\tau (ug_x+vg_y+g_t).
\end{align*}
With the higher order truncations, the Burnett and super-Burnett
equations can be obtained \cite{GKS-Xu1, GKS-Xu2}.

\begin{figure}[!h]
\centering
\includegraphics[width=0.4\textwidth]{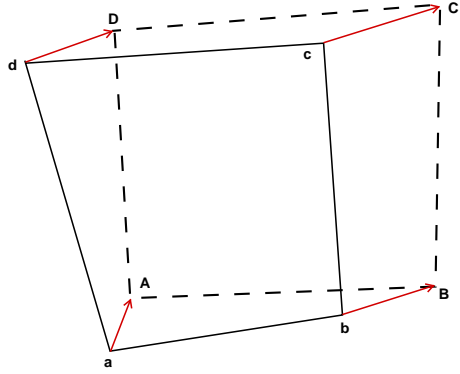}
\caption{\label{moving-1}Schematic for moving control volume:
$\Omega(t)$ moves from $[abcd]$ at $t^n$ to $[ABCD]$ at $t^{n+1}$.}
\end{figure}

In this paper, the high-order ALE gas-kinetic scheme will be
constructed in the moving-mesh framework. As shown in
Fig.\ref{moving-1}, a quadrilateral cell $\Omega(t)$ moves from
$[abcd]$ at $t^n$ to $[ABCD]$ at $t^{n+1}$ with grid velocity
$\boldsymbol{U}^{g}=(U^{g}, V^{g})$, which varies along each
segment. The boundary is given by
\begin{equation*}
\partial\Omega_i(t)=\bigcup_m\Gamma_{im}(t),
\end{equation*}
where $\Gamma_{im}(t)$ is the line segment of boundary. For
simplicity, the mesh velocity is set to be constant during a time
interval, and the segments keep straight in a time step. Standing on
the moving reference, the BGK equation Eq.\eqref{bgk} becomes
\begin{equation}\label{bgk2}
f_t+(u-U^g)f_x+(v-V^g)f_y =\frac{g-f}{\tau}.
\end{equation}
Taking moments of the kinetic equation Eq.\eqref{bgk2} and
integrating with respect to space, the finite volume scheme can be
expressed as
\begin{align}\label{finite}
\frac{\text{d}(|\Omega_i|W_{i})}{\text{d}t}=-\sum_mF_{im}(t),
\end{align}
where $W_i$ is the cell averaged conservative value of $\Omega_i$,
$|\Omega_i|$ is the area of $\Omega_i$ and $F_{im}(t)$ is the time
dependent fluxes across cell interface $\Gamma_{im}$, which is a
line integral over $\Gamma_{im}$
\begin{align*}
F_{im}(t)=&\int_{\Gamma_{im}}(\int\psi
f(\boldsymbol{x}(s),t,\boldsymbol{u},\xi)
(\boldsymbol{u}-\boldsymbol{U}^g) \cdot
\boldsymbol{n}_{im}\text{d}\Xi)\text{d}s,
\end{align*}
where $f(\boldsymbol{x}(s),t,\boldsymbol{u},\xi)$ is the gas
distribution function in the global coordinate and
$\boldsymbol{n}_{im}=(L_{im}, M_{im})$ is the outer normal direction
of $\Gamma_{im}$. To simplify the calculation, the Gaussian
quadrature is used for flux calculation
\begin{align*}
F_{im}(t)=|\Gamma_{im}|\sum_p\omega_{i,m_p}F_{i,m_p}(t).
\end{align*}
The flux at the Gaussian quadrature points is defined as
\begin{align*}
F_{i,m_p}(t)=\left(
\begin{array}{c}
F^{\rho}_{i,m_p}  \\
F^{m}_{i,m_p}  \\
F^{n}_{i,m_p}  \\
F^{E}_{i,m_p}  \\
\end{array}\right)=\int\psi f(\boldsymbol{x}_{i,m_p},t,\boldsymbol{u},\xi)
(\boldsymbol{u}-\boldsymbol{U}_{i,m_p}^g) \cdot
\boldsymbol{n}_{im}\text{d}\Xi,
\end{align*}
where $\boldsymbol{x}_{i,m_p}$ is the Gaussian quadrature point of
$\Gamma_{im}$, $\omega_{i,m_p}$ is the quadrature weight and
$\boldsymbol{U}_{i,m_p}^g$ is the grid velocity at quadrature point.
With the Gaussian quadrature,  the high-order spatial accuracy is
achieved, and the grid velocity variation along a cell interface is
considered, in which the translation, rotation and deformation are
taken into account. With identified normal direction, the
relative particle velocity at the Gaussian point in the local
coordinate is defined as
\begin{align}\label{velocity}
\begin{cases}
\widetilde{u}=(u-U_{i,m_p}^g)L_{im}+(v-V_{i,m_p}^g)M_{im},\\
\widetilde{v}=-(u-U_{i,m_p}^g)M_{im}+(v-V_{i,m_p}^g)L_{im}.
\end{cases}
\end{align}
In the actual computation, the spatial reconstruction, which will be
presented in the later section, is performed in a local moving
coordinate. With the reconstructed variables, the gas distribution
function is obtained at Gaussian quadrature point. The numerical
flux can be obtained by taking moments of it, and the component-wise
form can be written as
\begin{align*}
\widetilde{F}_{i,m_p}(t)=\left(
\begin{array}{c}
F^{\widetilde{\rho}}_{i,m_p}  \\
F^{\widetilde{m}}_{i,m_p}  \\
F^{\widetilde{n}}_{i,m_p}  \\
F^{\widetilde{E}}_{i,m_p}  \\
\end{array}\right)=
\int\widetilde{u}\left(
\begin{array}{c}
1\\
\widetilde{u}  \\
\widetilde{v}  \\
\frac{1}{2}(\widetilde{u}^2+\widetilde{v}^2+\xi^2) \\
\end{array}\right)
f(\boldsymbol{x}_{i,m_p},t,\widetilde{\boldsymbol{u}},\xi)\text{d}\widetilde{\Xi},
\end{align*}
where $f(\boldsymbol{x}_{i,m_p},t,\widetilde{\boldsymbol{u}},\xi)$
is the gas distribution function in the local coordinate.  According
to Eq.\eqref{velocity}, the fluxes in the inertia frame of reference
can be obtained as a combination of the fluxes in the moving frame
of reference
\begin{align}\label{flux-trans}
\left\{\begin{aligned}
F^{\rho}_{i,m_p}=&F^{\widetilde{\rho}}_{i,m_p},\\
F^{m}_{i,m_p}=&\big(U_{i,m_p}^gF^{\widetilde{\rho}}_{i,m_p}+M_{im}F^{\widetilde{m}}_{i,m_p}+L_{im}F^{\widetilde{n}}_{i,m_p}\big),\\
F^{n}_{i,m_p}=&\big(V_{i,m_p}^gF^{\widetilde{\rho}}_{i,m_p}-L_{im}F^{\widetilde{m}}_{i,m_p}+M_{im}F^{\widetilde{n}}_{i,m_p}\big),\\
F^{E}_{i,m_p}=&\big(F^{\widetilde{E}}_{i,m_p}+\frac{1}{2}((U_{i,m_p}^g)^2+(V_{i,m_p}^g)^2)F^{\widetilde{\rho}}_{i,m_p}\\
&+(M_{im}U_{i,m_p}^g-L_{im}V_{i,m_p}^g)F^{\widetilde{m}}_{i,m_p}+(L_{im}U_{i,m_p}^g+M_{im}V_{i,m_p}^g)F^{\widetilde{n}}_{i,m_p}\big).
\end{aligned} \right.
\end{align}

To construct the gas distribution function in the local
coordinate, the integral solution of the BGK equation at a cell
interface is used
\begin{equation}\label{integral1}
f(\boldsymbol{x}_{i,m_p},t,\boldsymbol{u},\xi)=\frac{1}{\tau}\int_0^t g(x',y',t',u,v,\xi)e^{-(t-t')/\tau}\text{d}t'\\
+e^{-t/\tau}f_0(-ut,-vt,u,v,\xi),
\end{equation}
where $\widetilde{\boldsymbol{u}}$ is simply denoted as
$\boldsymbol{u}$, $\boldsymbol{x}_{i,m_p}=(x_{i,m_p},y_{i,m_p})$ is
the location of Gaussian quadrature point, $x_{i,m_p}=x'+u(t-t'),
y_{i,m_p}=y'+v(t-t')$ are the trajectory of particles,  $f_0$ is the
initial gas distribution function, and $g$ is the corresponding
equilibrium state. According to Eq.\eqref{integral1}, the time
dependent gas distribution function
$f(\boldsymbol{x}_{i,m_p},t,\boldsymbol{u},\xi)$ at the cell
interface can be expressed as \cite{GKS-Xu2,GKS-Xu1}
\begin{align}\label{flux}
f(\boldsymbol{x}_{i,m_p},t,\boldsymbol{u},\xi)=&(1-e^{-t/\tau})g_0+((t+\tau)e^{-t/\tau}-\tau)(\overline{a}_1u+\overline{a}_2v)g_0\nonumber\\
+&(t-\tau+\tau e^{-t/\tau}){\bar{A}} g_0\nonumber\\
+&e^{-t/\tau}g_r[1-(\tau+t)(a_{1r}u+a_{2r}v)-\tau A_r)]H(u)\nonumber\\
+&e^{-t/\tau}g_l[1-(\tau+t)(a_{1l}u+a_{2l}v)-\tau A_l)](1-H(u)),
\end{align}
where the coefficients in Eq.\eqref{flux} can be determined by the
reconstructed directional derivatives and compatibility condition
\begin{align*}
\displaystyle \langle a_{1}^{k}\rangle=\frac{\partial
W_{k}}{\partial \boldsymbol{n}}, \langle
a_{2}^{k}\rangle&=\frac{\partial W_{k}}{\partial \boldsymbol{\tau}},
\langle a_{1}^{k}u+a_{2}^{k}v+A^{k}\rangle=0,\\
\displaystyle \langle\overline{a}_1\rangle=\frac{\partial
\overline{W}}{\partial \boldsymbol{n}},
\langle\overline{a}_2\rangle&=\frac{\partial \overline{W}}{\partial
\boldsymbol{\tau}},
\langle\overline{a}_1u+\overline{a}_2v+\overline{A}\rangle=0,
\end{align*}
where  $k=l,r$, $\boldsymbol{n}$ and $\boldsymbol{\tau}$ are local
normal and tangential direction and $\langle...\rangle$ are the
moments of the equilibrium $g$ and defined by
\begin{align*}
\langle...\rangle=\int g (...)\psi \text{d}\Xi.
\end{align*}
The spatial derivatives will be obtained by the WENO reconstruction,
which will be given later. More details of the
gas-kinetic scheme can be found in \cite{GKS-Xu1}.

In the formulations above, the grid velocity $\boldsymbol{U}^g$ can
be arbitrary to control the volume evolution. The Eulerian governing
equation is obtained with $\boldsymbol{U}^g=0$, and the Lagrangian
form is obtained with $\boldsymbol{U}^g=\boldsymbol{U}$, where
$\boldsymbol{U}$ is the fluid velocity. In this paper, the
structured meshes are used in the moving mesh computation for
simplicity, while the unstructured reconstruction is used to deal
with the distorted meshes. The strategies of mesh velocity are given
as follows
\begin{enumerate}
\item The mesh velocity is specified directly. This is the simplest
type of mesh velocity and is mainly adopted in the accuracy tests in
this paper.
\item The mesh velocity can be given by the fluid velocity, and the
cell centered Lagrangian nodal solver is used \cite{mesh-vel-1}.
Denote $\boldsymbol{U}_p$ as the vertex fluid velocity of the
control volume, $\boldsymbol{U}_c$ as cell averaged fluid velocity,
$C(p)$ as the set of control volumes that share the common vertex
$p$,  the mesh velocity can be obtained by solving the following
linear system
\begin{equation*}
\sum_{c\in C(p)}\mathbb{M}_{pc}\boldsymbol{U}_p=\sum_{c\in
C(p)}l_{pc}p_c\boldsymbol{n}_{pc}+\mathbb{M}_{pc}\boldsymbol{U}_c,
\end{equation*}
where
\begin{align*}
\mathbb{M}_{pc}=\rho_c&a_c[l_{pc}^-(\boldsymbol{n}_{pc}^-\otimes\boldsymbol{n}_{pc}^-)+l_{pc}^+(\boldsymbol{n}_{pc}^+\otimes\boldsymbol{n}_{pc}^+)],\\
&l_{pc}\boldsymbol{n}_{pc}=l_{pc}^-\boldsymbol{n}_{pc}^-+l_{pc}^+\boldsymbol{n}_{pc}^+,
\end{align*}
where $pp^-$ and $pp^+$ are the line segments of $c$ that share the
vertex $p$. $l_{pc}^-, l_{pc}^+$ and $\boldsymbol{n}_{pc}^-,
\boldsymbol{n}_{pc}^+$ are the half lengthes and outer normal
directions of the two segments. The matrix $\mathbb{M}_{pc}$ is
symmetric positive definite. Therefore, the system will always admit
a unique solution.
\item
The mesh velocity can be determined by the variational approach
\cite{mesh-vel-2}, and the corresponding Euler-Lagrange equations
can be obtained from
\begin{align*}
(\omega x_\xi)_\xi&+(\omega x_\eta)_\eta=0,\\
(\omega y_\xi)_\xi&+(\omega y_\eta)_\eta=0,
\end{align*}
where $(x,y)$ and $(\xi,\eta)$ denote the physical and computational
coordinates, $\omega$ is the monitor function and can be chosen as a
function of the flow variables, such as the density, velocity,
pressure, or their gradients. In the numerical tests, the monitor
function takes the form
\begin{equation*}
\omega=\sqrt{1+\alpha|\nabla \rho|^2},
\end{equation*}
if no special statement is given.
The mesh distribution in the physical space can be directly
generated \cite{mesh-vel-2}.
\end{enumerate}
In order to avoid mesh distortion, the new meshes obtained by the
Lagrangian method and adaption method are smoothed by the following
smooth procedure for the structured meshes
\begin{align*}
\widetilde{\boldsymbol{x}}^{n+1}_{ij}=&(4\boldsymbol{x}^{n+1}_{ij}+\boldsymbol{x}^{n+1}_{i-1,j}+\boldsymbol{x}^{n+1}_{i+1,j}+\boldsymbol{x}^{n+1}_{i,j-1}+\boldsymbol{x}^{n+1}_{i,j+1})/8,
\end{align*}
where $\boldsymbol{x}_{ij}$ is the coordinate of the grid points.
This procedure is conducted for some certain steps. The velocity at
grid points $\textbf{U}_{ij}^g=(U_{ij}^g, V_{ij}^g)$ can be
determined as follows
\begin{align*}
\boldsymbol{U}_{ij}^g=\frac{\widetilde{\boldsymbol{x}}^{n+1}_{ij}-\boldsymbol{x}^n_{ij}}{\Delta
t}.
\end{align*}
The mesh velocity $\boldsymbol{U}_{i,j_p}^g$ and
$\boldsymbol{U}_{i_p,j}^g$ at Gaussian quadrature points can be
obtained by the linear interpolation of mesh velocity at the grid
points.

\section{Two-stage temporal discretization}
A two-stage fourth-order time-accurate discretization was developed
for Lax-Wendroff flow solvers, such as the generalized Riemann
problem (GRP) solver and the gas-kinetic scheme \cite{GRP-GKS}.
Consider the following time-dependent equation
\begin{align}\label{pde}
\frac{\partial \boldsymbol{w}}{\partial t}=\mathcal
{L}(\boldsymbol{w}),
\end{align}
with the initial condition at $t_n$, i.e.,
\begin{align*}
\boldsymbol{w}(t=t_n)=\boldsymbol{w}^n,
\end{align*}
where $\mathcal {L}$ is an operator for spatial derivative of flux.
Introducing an intermediate state at $t^*=t_n+\Delta t/2$,
\begin{align}\label{step1}
\boldsymbol{w}^*=\boldsymbol{w}^n+\frac{1}{2}\Delta t\mathcal
{L}(\boldsymbol{w}^n)+\frac{1}{8}\Delta t^2\frac{\partial}{\partial
t}\mathcal{L}(\boldsymbol{w}^n),
\end{align}
 the state $\boldsymbol{w}$ can be updated with the following
formula
\begin{align}\label{step2}
\boldsymbol{w}^{n+1}=\boldsymbol{w}^n+\Delta t\mathcal
{L}(\boldsymbol{w}^n)+\frac{1}{6}\Delta
t^2\big(\frac{\partial}{\partial
t}\mathcal{L}(\boldsymbol{w}^n)+2\frac{\partial}{\partial
t}\mathcal{L}(\boldsymbol{w}^*)\big).
\end{align}
It can be proved that Eq.\eqref{step1} and Eq.\eqref{step2} provide
a fourth-order time accurate solution for Eq.\eqref{pde} at $t=t_n
+\Delta t$.  The details of proof can be found in \cite{GRP-high-1}.

\begin{figure}[!h]
\centering
\includegraphics[width=0.4\textwidth]{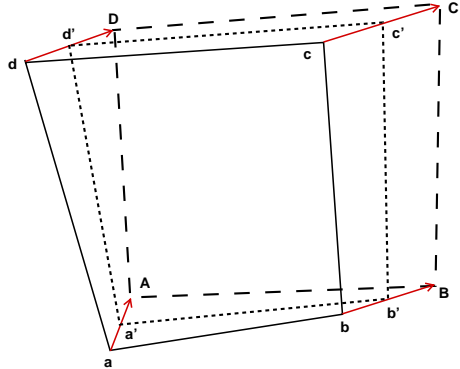}
\caption{\label{moving-2}Schematic for a moving control volume:
$\Omega(t^*)=[a'b'c'd']$ is the intermediate one at $t^*=t_n+\Delta
t/2$ and $a',b',c',d'$ are the mid-points of $aA, bB,cC,dD$.}
\end{figure}

To achieve the high-order spatial and temporal accuracy, the
two-stage temporal discretization can be extended to the moving mesh
computation Eq.\eqref{finite}, where the operator $\mathcal{L}$ is
denoted as
\begin{align*}
\mathcal{L}(\Omega_i,W_{i})=-\sum_mF_{im}(t).
\end{align*}
As shown in Fig.\ref{moving-2}, the coordinate of the local moving
reference and the length of cell interface are both time-dependent,
which is critical for accuracy and geometry conservation law. In the
two-stage method, the variables are defined and the reconstruction
is performed at $t_n$ and $t^*=t_n+\Delta t/2$, respectively. Due to
the constant velocity for each vertex, the intermediate control
volume at $t^*$ moves to $\Omega(t^*)=[a'b'c'd']$ exactly, where
$a',b',c',d'$ are the mid-points of $aA, bB,cC,dD$. To implement the
two-stage method, the WENO reconstructions is performed with the
re-distributed meshes at $t_n$ and $t^*$. With the reconstructed
variables, $\mathcal {L}(W)$ and temporal derivative
$\partial_t\mathcal {L}(W)$ for $\Omega_i$ at $t_n$ can be written
as
\begin{align*}
\mathcal{L}(\Omega_i^n,W_{i}^n)&=-\sum_m\sum_p\omega_{m_p}|\Gamma_{im}^n|\cdot
F_{i,m_p}^n,\\
\partial_t\mathcal {L}(\Omega_i^n,W_{i}^n)&=-\sum_m\sum_p\omega_{m_p}|\Gamma_{im}^n|\cdot
\partial_tF_{i,m_p}^n,
\end{align*}
where the coefficient $F_{i,m_p}^n$ and $\partial_tF_{i,m_p}^n$ can
be given by the linear combination of  $\widetilde{F}_{i,m_p}^n$ and
$\partial_t\widetilde{F}_{i,m_p}^n$, which are the coefficients of
linear approximation of the time dependent flux in the local
coordinate
\begin{align}\label{expansion-1}
\widetilde{F}_{i,m_p}(t)=\widetilde{F}_{i,m_p}^n+ \partial_t
\widetilde{F}_{i,m_p}^n(t-t_n).
\end{align}
To implement two-stage method, the following notation is introduced
\begin{align*}
\widehat{\mathbb{F}}(\boldsymbol{x}_{i,m_p},\delta)
=\int_{t_n}^{t_n+\delta}\widetilde{F}_{i,m_p}(t)\text{d}t=\int_{t_n}^{t_n+\delta}\int
\widetilde{u}\widetilde{\psi}
f(\boldsymbol{x}_{i,m_p},t,\widetilde{\boldsymbol{u}},\xi)\text{d}\Xi\text{d}t.
\end{align*}
Integrating Eq.\eqref{expansion-1} over $[t_n, t_n+\Delta t/2]$ and
$[t_n, t_n+\Delta t]$, we have the following two equations
\begin{align*}
\widetilde{F}_{i,m_p}^n\Delta t&+\frac{1}{2}\partial_t \widetilde{F}_{i,m_p}^n\Delta t^2 =\widehat{\mathbb{F}}(\boldsymbol{x}_{i,m_p},\Delta t) , \\
\frac{1}{2}\widetilde{F}_{i,m_p}^n\Delta t&+\frac{1}{8}\partial_t
\widetilde{F}_{i,m_p}^n\Delta t^2
=\widehat{\mathbb{F}}(\boldsymbol{x}_{i,m_p}, \Delta t/2).
\end{align*}
The coefficients $\widetilde{F}_{i,m_p}^n$ and $\partial_t
\widetilde{F}_{i,m_p}^n$ can be determined by solving the linear
system.  Similarly, $\mathcal {L}(\Omega_i^*,W_{i}^*)$ and temporal
derivative $\partial_t\mathcal {L}(\Omega_i^*,W_{i}^*)$ at the
intermediate state can be constructed as well. According to
Eq.\eqref{flux-trans}, the coefficients $F_{i,m_p}^n$ and
$\partial_t F_{i,m_p}^n$ in the global coordinate can be obtained.
More details of the two-stage fourth-order scheme can be found in
\cite{GRP-high-1,GKS-high-1}.

\section{Unstructured WENO reconstruction}
During the moving-mesh procedure, the mesh becomes distorted and the
development of robust high-order scheme on a low quality mesh is demanding.
Recently, a third-order WENO reconstruction is
constructed on quadrilateral and triangular meshes \cite{un-WENO3},
in which high-order accuracy is achieved and robustness is improved.
In this section, the third-order WENO reconstruction is extended to
the framework of gas-kinetic scheme. For the moving-mesh
computation, the spatial reconstruction is performed in the local
moving coordinate at each Gaussian quadrature point, where the
variation of mesh velocity is considered along a cell interface.

\begin{figure}[!htb]
\centering
\includegraphics[width=0.45\textwidth]{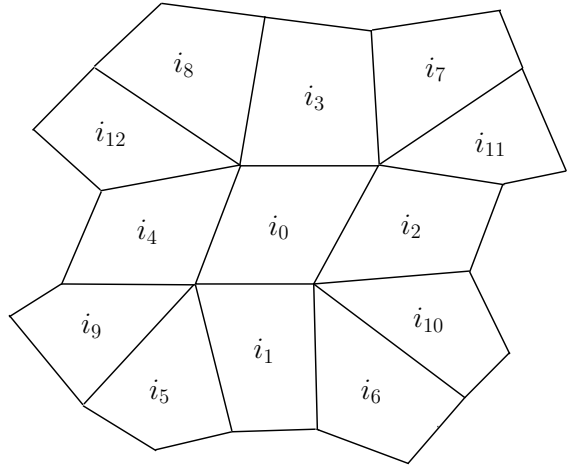}
\caption{\label{quadstencil} Stencils for third-order schemes on
quadrilateral meshes.}
\end{figure}

In this section, the reconstruction on quadrilateral mesh is
reviewed briefly and more details can be found in \cite{un-WENO3}.
To improve the robustness of WENO scheme, a general selection of
stencils is given as shown in Fig.\ref{quadstencil}.  To achieve the
third-order accuracy, the following quadratic polynomial $P^2(x,y)$
can be constructed
\begin{equation}\label{polynomial-eta}
P^2(x,y)=W_{i_0}+\sum_{k=1}^5a_kp^k(x,y),
\end{equation}
where $W_{i_0}$ is the cell average value of $W(x,y)$ over cell
$\Omega_{i_0}$ and $p^k(x,y), k=1,...,5$ are basis functions, which
are given as follows
\begin{align}\label{base}
\begin{cases}
\displaystyle p^1(x,y)=x-\frac{1}{\left| \Omega_{i_0} \right|}\displaystyle\iint_{\Omega_{i_0}}x \text{d}x\text{d}y, \\
\displaystyle p^2(x,y)=y-\frac{1}{\left| \Omega_{i_0} \right|}\displaystyle\iint_{\Omega_{i_0}}y \text{d}x\text{d}y, \\
\displaystyle p^3(x,y)=x^2-\frac{1}{\left| \Omega_{i_0} \right|}\displaystyle\iint_{\Omega_{i_0}}x^2 \text{d}x\text{d}y, \\
\displaystyle p^4(x,y)=y^2-\frac{1}{\left| \Omega_{i_0} \right|}\displaystyle\iint_{\Omega_{i_0}}y^2 \text{d}x\text{d}y, \\
\displaystyle p^5(x,y)=xy-\frac{1}{\left| \Omega_{i_0}
\right|}\displaystyle\iint_{\Omega_{i_0}}xy dxdy.
\end{cases}
\end{align}
With the following constraint over the cell $\Omega_{i_j}, i_j\in
S=\{i_0,i_1,...,i_{12}\}$,
\begin{align*}
\frac{1}{\left| \Omega_{i_j}
\right|}\int_{\Omega_{i_j}}P^2(x,y)\text{d}x\text{d}y=W_{i_j},
\end{align*}
the coefficients $\textbf{a}=(a_1,...,a_5)$ in
Eq.\eqref{polynomial-eta} can be fully given, where $W_{i_j}$ is the
cell averaged variables over cell $\Omega_{i_j}$ .

Similar with the standard WENO reconstruction \cite{Hu-Shu}, twelve
sub-stencils $S_{j}, j=1,...,12$ are selected from the large stencil
given in Fig.\ref{quadstencil}
\begin{align*}
P_{1}^1 ~&\text{on}~ S_1=\{i_0,i_1,i_2\}, ~~~P_{2}^1 ~\text{on}~ S_2=\{i_0,i_2,i_3\}, ~~~P_{3}^1 ~\text{on}~ S_3=\{i_0,i_3,i_4\},\\
P_{4}^1 ~&\text{on}~ S_4=\{i_0,i_4,i_1\}, ~~~P_{5}^1 ~\text{on}~ S_5=\{i_0,i_1,i_5\}, ~~~P_{6}^1 ~\text{on}~ S_6=\{i_0,i_1,i_6\},\\
P_{7}^1 ~&\text{on}~ S_7=\{i_0,i_3,i_7\}, ~~~P_{8}^1 ~\text{on}~ S_8=\{i_0,i_3,i_8\}, ~~~P_{9}^1 ~\text{on}~ S_9=\{i_0,i_4,i_9\},\\
P_{10}^1 ~&\text{on}~ S_{10}=\{i_0,i_2,i_{10}\}, P_{11}^1
~\text{on}~ S_{11}=\{i_0,i_2,i_{11}\}, P_{12}^1 ~\text{on}~
S_{12}=\{i_0,i_4,i_{12}\}.
\end{align*}
Twelve candidate linear polynomials $P_{j}^1$ corresponding to the
candidate sub-stencils can be constructed as follows
\begin{align*}
\frac{1}{\left|
\Omega_{S_{j,k}}\right|}\int_{\Omega_{S_{j,k}}}P^1_j(x,y)\text{d}x\text{d}y=W_{S_{j,k}},
~k=0,1,2,
\end{align*}
where $\Omega_{S_{j,k}}$ is the $(k+1)$-th cell in the sub-stencils
$S_{j}$, and $W_{S_{j,k}}$ is the cell averaged value over cell
$\Omega_{S_{j,k}}$.

With point value of the quadratic polynomial  $P^2(x,y)$ and linear
polynomials $P_{j}^1$ at the Gaussian quadrature point $(x_G,y_G)$,
a standard procedure of WENO scheme is adopted to obtain the
linear weights $\gamma_j$ \cite{Hu-Shu}. However, in the traditional
WENO reconstruction, the very large linear weights and negative
weights appear with the lower mesh quality. In order to improve the
robustness of WENO schemes, an optimization approach is given to
deal with the very large linear weights. The weighting parameters is
introduced for each cell, such that the ill cell contributes little
to the quadratic polynomial in Eq.\eqref{polynomial-eta}. For
quadrilateral meshes, the weighting parameters are defined for each
cell
\begin{equation*}
\begin{cases}
d_j=1, & ~~j=1,...,4, \\
d_j=\displaystyle\frac{1}{\max(1,\left| \gamma_j\right|)}, &
~~j=5,...,12.
\end{cases}
\end{equation*}
The optimized coefficients $\textbf{a}=(a_1,...,a_5)$ of the
quadratic polynomial in Eq.\eqref{polynomial-eta} are given by the
following weighted linear system
\begin{equation*}
\sum_{k=1}^5 d_j\cdot A_{jk}a_k=d_j\cdot (W_{i_j}-W_{i_0}).
\end{equation*}
With the procedure above, the maximum $\gamma_j$ becomes the order
$O(1)$ \cite{un-WENO3}.

To deal with the negative linear weights, the splitting technique
\cite{splitting-weights} is considered for the optimized approach as
follows
\begin{equation*}
\widetilde{\gamma}_j^{+}=\frac{1}{2}(\gamma_j + \theta
\left|\gamma_j\right|),
~~\widetilde{\gamma}_j^{-}=\widetilde{\gamma}_j^{+}-\gamma_j,
\end{equation*}
where $\theta=3$ is taken in numerical tests. The scaled
non-negative linear weights $\gamma_j^{\pm}$ and nonlinear weights
$\delta_{j}^{\pm}$ are given by
\begin{align*}
\gamma_j^{\pm}=\frac{\widetilde{\gamma}_j^{\pm}}{\sigma^{\pm}},&~\sigma^{\pm}=\sum_{j}\widetilde{\gamma}_j^{\pm},\\
\delta_{j}^{\pm}=\frac{\alpha_{j}^{\pm}}{\sum_{l=1}\alpha_{l}^{\pm}},&~~\alpha_{j}^{\pm}=\frac{\gamma_{j}^{\pm}}{(\widetilde{\beta}_j^{\pm}+\epsilon)^{2}},
\end{align*}
where $\epsilon$ is a small positive number,
$\widetilde{\beta_j}^{\pm}$ is a new smooth indicator defined on
unstructured meshes
\begin{equation*}
\widetilde{\beta_j}^{\pm}=\beta_j\big(1+\gamma_j^{\pm}\beta_j+(\gamma_j^{\pm}
\beta_j)^2 \big),
\end{equation*}
where $\beta_{j}$ is defined as
\begin{equation*}
\beta_j=\sum_{|\alpha|=1}^{K}|\Omega|^{|\alpha|-1}\iint_{\Omega}\big(D^{\alpha}P_j^1(x,y)\big)^2\text{d}x\text{d}y,
\end{equation*}
$\alpha$ is a multi-index, and $D$ is the derivative operator. With
the new smooth indicator, the accuracy keeps the original order in
smooth regions with $IS=h^2(1+O(h))$. The final reconstructed value
at the Gaussian quadrature points can be written as
\begin{equation*}
R(x_G,y_G)=\sum_{j}(\delta_j^{+}P^1_j(x_G,y_G)-\delta_j^{-}P^1_j(x_G,y_G)).
\end{equation*}

Compared with the numerical scheme based on the Riemann solvers, the
reconstruction corresponding equilibrium state is also presented.
With the reconstructed variables $W_l$ and $W_r$ at both sides of
the cell interface, the variables at the cell interface can be
determined by the compatibility condition Eq.\eqref{compatibility}
as follows
\begin{align*}
\int\psi g_{G}\text{d}\Xi=W_G=\int_{u>0}\psi
g_{l}\text{d}\Xi+\int_{u<0}\psi g_{r}\text{d}\Xi,
\end{align*}
where  $g_{l}$ and $g_{r}$ are the equilibrium states corresponding
to $W_l$ and $W_r$ respectively. To third-order accuracy, the Taylor
expansion corresponding to equilibrium state at the center of cell
interface is expressed as
\begin{align*}
&\overline{W}(x,y)=W_G+\overline{W}_{ \boldsymbol{n}}(x-x_G)+\overline{W}_{\boldsymbol{\tau}}(y-y_G)\nonumber \\
+\frac{1}{2}\overline{W}_{ \boldsymbol{n}
\boldsymbol{n}}&(x-x_G)^2+\overline{W}_{ \boldsymbol{n}
\boldsymbol{\tau}}(x-x_G)(y-y_G)+\frac{1}{2}\overline{W}_{\boldsymbol{\tau}\boldsymbol{\tau}}(y-y_G)^2,
\end{align*}
where $\boldsymbol{n}, \boldsymbol{\tau}$ are local normal and
tangential directions. With the following constrains
\begin{align*}
\frac{1}{\left| \Omega_{i_j}
\right|}\int_{\Omega_{i_j}}\overline{W}(x,y)\text{d}x\text{d}y=W_{i_j},
i_j\in T=\{i_0,i_1,...,i_{7}\},
\end{align*}
the spatial derivatives can be obtained by the least square methods,
where the stencils for equilibrium state reconstruction is given in
Fig.\ref{quadstencil}. For most mesh generation method, there are at
least five elements in the above stencil, which guarantees the least
square method solvable.

\begin{figure}[!htb]
\centering
\includegraphics[width=0.375\textwidth]{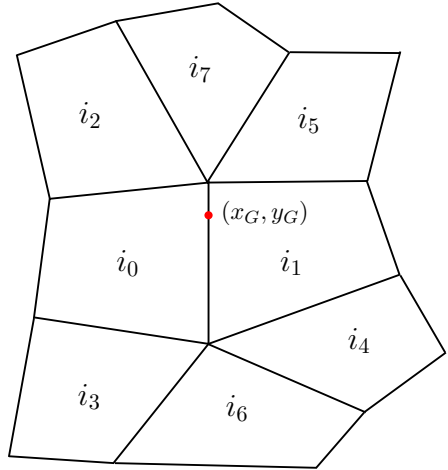}
\caption{\label{quadstencil} Stencils for equilibrium state of
third-order scheme on quadrilateral meshes.}
\end{figure}

\begin{figure}[!htb]
\centering
\includegraphics[width=0.55\textwidth]{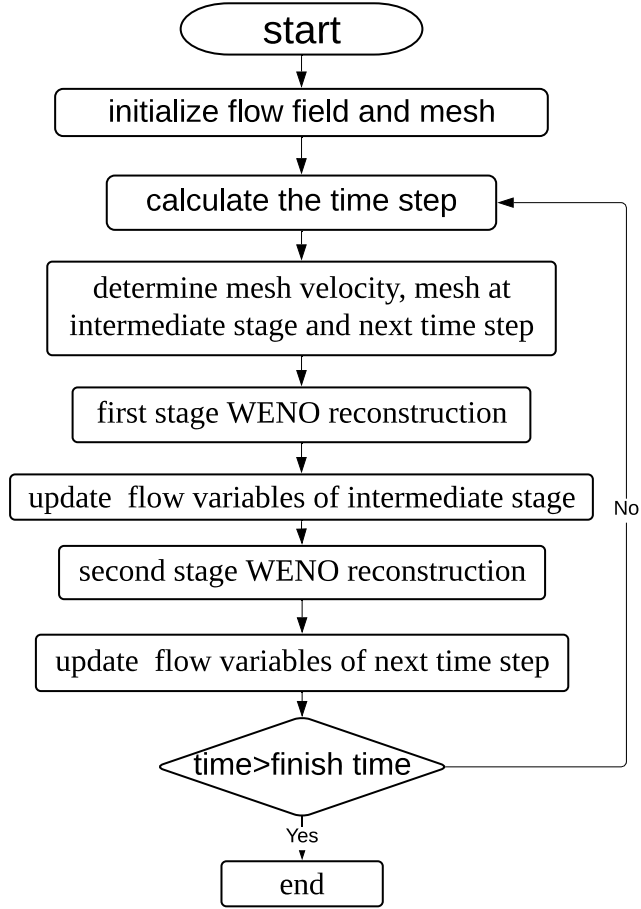}
\caption{\label{chart} Flow chart of high-order ALE gas-kinetic
scheme.}
\end{figure}

Based on the reconstruction corresponding to the equilibrium and
non-equilibrium parts, the reconstructed variables and the gas
distribution in the local moving coordinate can be obtained, and the
flow variables can be updated by the two-stage framework.

\section{Numerical tests}
In this section, numerical tests for both inviscid and viscous flows
will be presented to validate our numerical scheme. For the inviscid
flow, the collision time $\tau$ takes
\begin{align*}
\tau=\epsilon \Delta t+C\displaystyle|\frac{p_l-p_r}{p_l+p_r}|\Delta
t,
\end{align*}
where $\epsilon=0.01$ and $C=1$. For the viscous flow, we have
\begin{align*}
\tau=\frac{\mu}{p}+C \displaystyle|\frac{p_l-p_r}{p_l+p_r}|\Delta t,
\end{align*}
where $p_l$ and $p_r$ denote the pressure on the left and right
sides of the cell interface, $\mu$ is the dynamic viscous
coefficient, and $p$ is the pressure at the cell interface. In
smooth flow regions, it will reduce to $\tau=\mu/p$. The ratio of
specific heats takes $\gamma=1.4$.

To validate the WENO based gas-kinetic scheme, the accuracy test on
the unstructured quadrilateral meshes is presented with the
stationary mesh first. For the moving-mesh computation, the structured
meshes are used to avoid the distortion of computational mesh, while
the procedure of reconstruction is still based on the unstructured
code. The flow chart for the high-order ALE scheme is presented in
Fig.\ref{chart}.

\begin{figure}[!htb]
\centering
\includegraphics[width=0.475\textwidth]{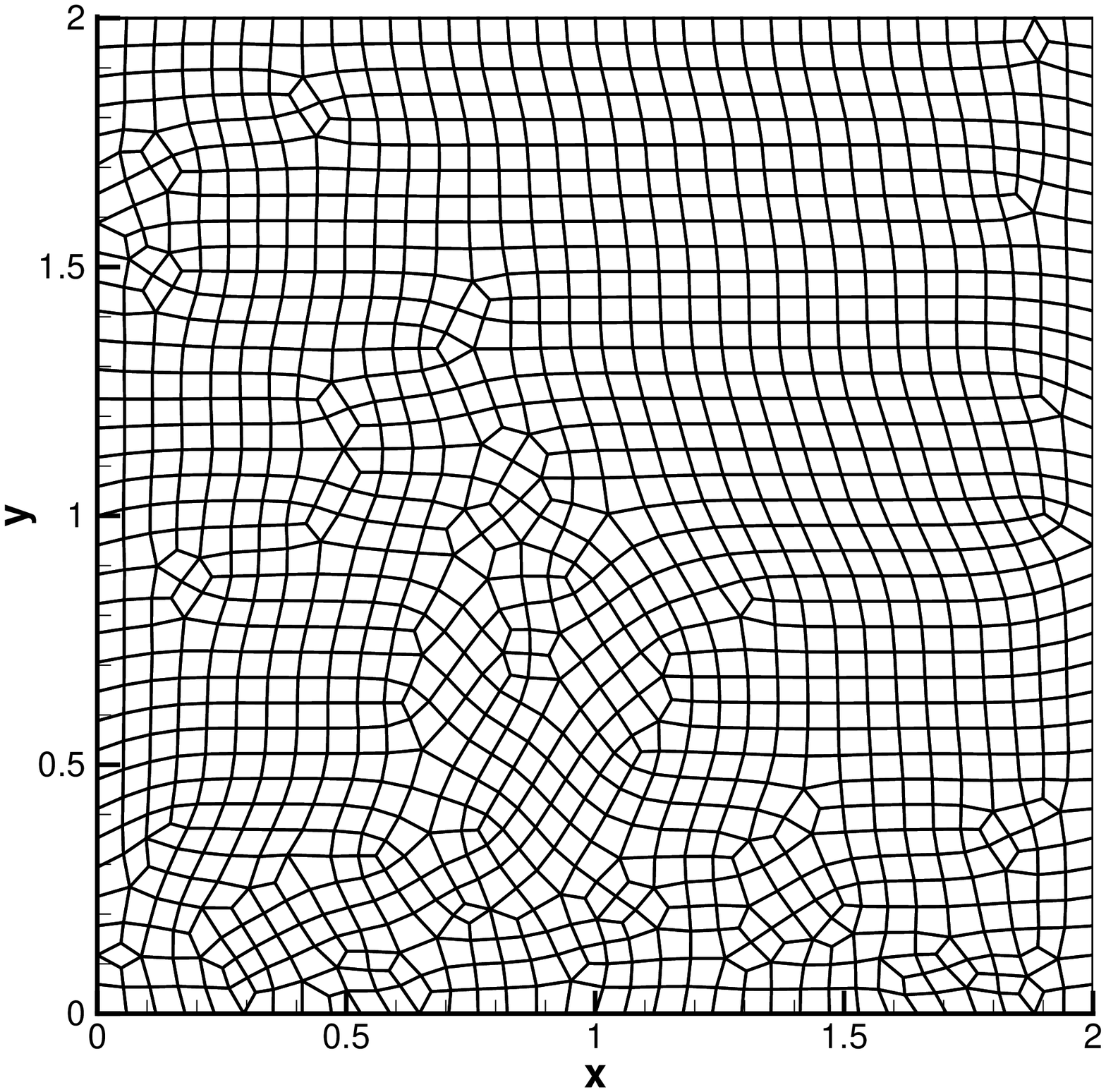}
\includegraphics[width=0.475\textwidth]{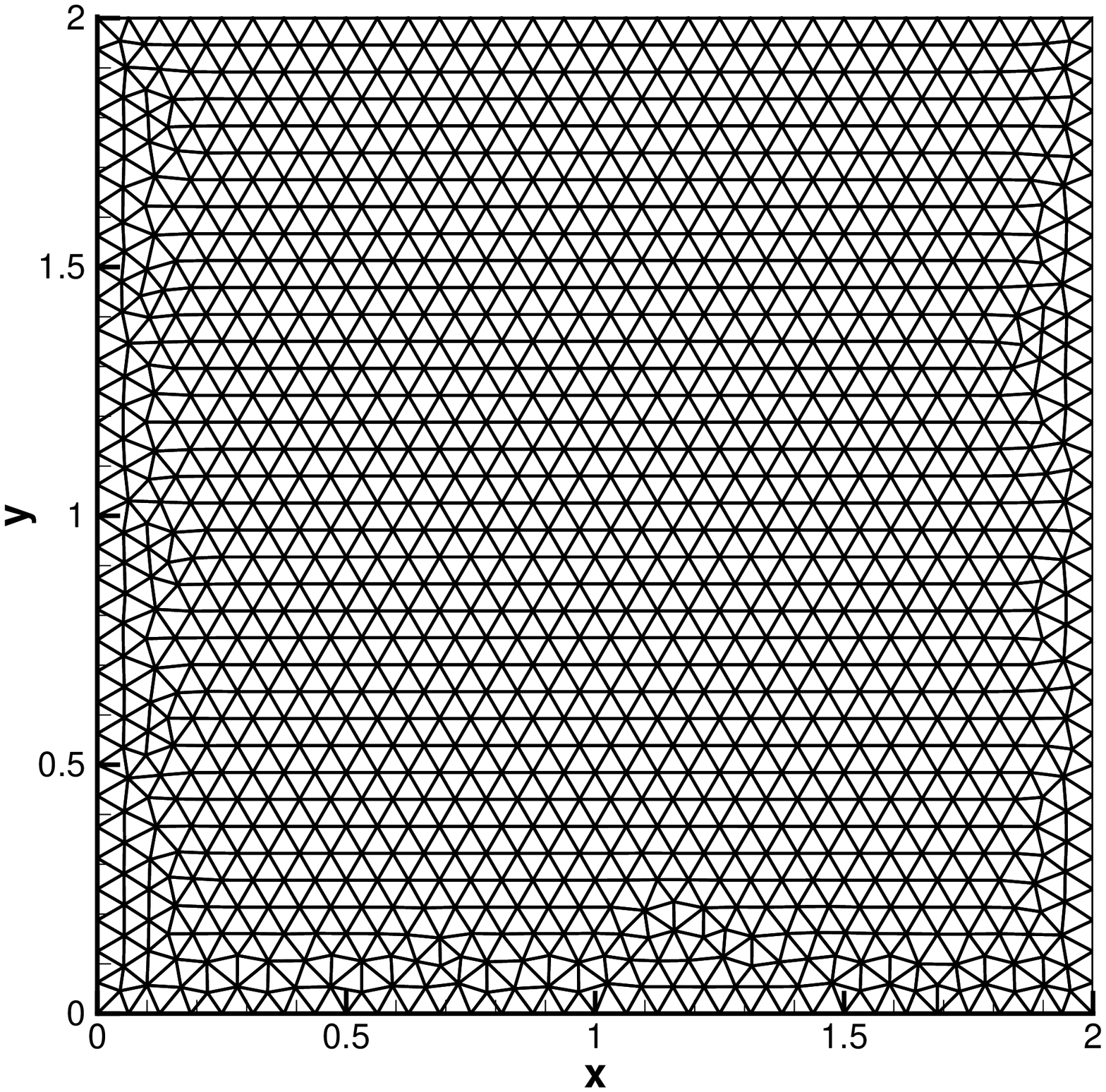}
\caption{\label{Mesh-Accuracy-1} Accuracy test: the unstructured
quadrilateral and triangular meshes with cell size $h=1/16$.}
\end{figure}

\begin{table}[!h]
\begin{center}
\def\temptablewidth{0.75\textwidth}
{\rule{\temptablewidth}{1.0pt}}
\begin{tabular*}{\temptablewidth}{@{\extracolsep{\fill}}c|cc|cc}
mesh    & $L^1$ error  &  Order    & $L^2$ error  &  Order   \\
\hline
$1/8$  &  1.3752E-02  &  ~      &  1.5288E-02  &  ~        \\
$1/16$ &  1.4112E-03  &  3.28   &  1.5684E-03  &  3.28   \\
$1/32$ &  2.0183E-04  &  2.80   &  2.2400E-04  &  2.80   \\
$1/64$ &  2.3544E-05  &  3.09   &  2.6098E-05  &  3.10   \\
\end{tabular*}
{\rule{\temptablewidth}{1.0pt}}
\end{center}
\vspace{-5mm}\caption{\label{accuracy-quad1} Accuracy test: WENO-3rd
with linear weights on quadrilateral meshes.}
\begin{center}
\def\temptablewidth{0.75\textwidth}
{\rule{\temptablewidth}{1.0pt}}
\begin{tabular*}{\temptablewidth}{@{\extracolsep{\fill}}c|cc|cc}
mesh    & $L^1$ error  &  Order    & $L^2$ error  &  Order   \\
\hline
$1/8$  &  6.0428E-02  &   ~     &  6.9740E-02  &   ~       \\
$1/16$ &  1.2965E-02  &   2.22  &  1.5577E-02  &   2.16  \\
$1/32$ &  2.2794E-03  &   2.50  &  2.7735E-03  &   2.48  \\
$1/64$ &  1.1090E-04  &   4.36  &  1.4425E-04  &   4.26
\end{tabular*}
{\rule{\temptablewidth}{1.0pt}}
\end{center}
\vspace{-5mm}\caption{\label{accuracy-quad2} Accuracy test: WENO-3rd
with non-linear weights on quadrilateral meshes.}
\begin{center}
\def\temptablewidth{0.75\textwidth}
{\rule{\temptablewidth}{1.0pt}}
\begin{tabular*}{\temptablewidth}{@{\extracolsep{\fill}}c|cc|cc}
mesh    & $L^1$ error  &  Order    & $L^2$ error  &  Order   \\
\hline
$1/8$  &  4.8752E-03  &  ~        &  5.4117E-03  &  ~        \\
$1/16$ &  6.4576E-04  &  2.92     &  7.1650E-04  &  2.92     \\
$1/32$ &  8.2542E-05  &  2.97     &  9.1507E-05  &  2.97     \\
$1/64$ &  1.0308E-05  &  3.00     &  1.1427E-05  &  3.00     \\
\end{tabular*}
{\rule{\temptablewidth}{1.0pt}}
\end{center}
\vspace{-5mm}\caption{\label{accuracy-tri1} Accuracy test: WENO-3rd
with linear weights on triangular meshes.}
\begin{center}
\def\temptablewidth{0.75\textwidth}
{\rule{\temptablewidth}{1.0pt}}
\begin{tabular*}{\temptablewidth}{@{\extracolsep{\fill}}c|cc|cc}
mesh    & $L^1$ error  &  Order    & $L^2$ error  &  Order   \\
\hline
$1/8$  &  2.0211E-02  &   ~       &  2.5819E-02  &   ~     \\
$1/16$ &  4.6260E-03  &   2.13    &  5.4557E-03  &   2.24  \\
$1/32$ &  3.4571E-04  &   3.74    &  4.6289E-04  &   3.56  \\
$1/64$ &  1.7427E-05  &   4.31    &  2.4689E-05  &   4.23  \\
\end{tabular*}
{\rule{\temptablewidth}{1.0pt}}
\end{center}
\vspace{-5mm}\caption{\label{accuracy-tri2} Accuracy test: WENO-3rd
with non-linear weights on triangular meshes.}
\end{table}

\subsection{Accuracy tests}
The two-dimensional advection of density perturbation is used as the
accuracy test. The computational domain is $[0,2]\times[0,2]$ and
the initial conditions are given as follows
\begin{align*}
\rho_0(x,y)=&1+0.2\sin(\pi (x+y)),~p_0(x,y)=1,\\
&U_0(x,y)=1,~V_0(x,y)=1.
\end{align*}
The periodic boundary conditions are imposed at boundaries and the
exact solutions are
\begin{align*}
\rho(x,y,t)=&1+0.2\sin(\pi(x+y-t)),~p(x,y,t)=1,\\
&U(x,y,t)=1,~V(x,y,t)=1.
\end{align*}
The accuracy of the third-order WENO-based gas-kinetic scheme on
stationary unstructured quadrilateral and triangular meshes are
tested, and the meshes with cell size $h=1/16$ are given in
Fig.\ref{Mesh-Accuracy-1} as reference. The $L^1$ and $L^2$ errors
and convergence orders for current scheme are presented in
Tab.\ref{accuracy-quad1} and Tab.\ref{accuracy-quad2} for
quadrilateral meshes, and in Tab.\ref{accuracy-tri1} and
Tab.\ref{accuracy-tri2} for triangular meshes.  The expected order
of accuracy is achieved for the current scheme.

\begin{figure}[!htb]
\centering
\includegraphics[width=0.475\textwidth]{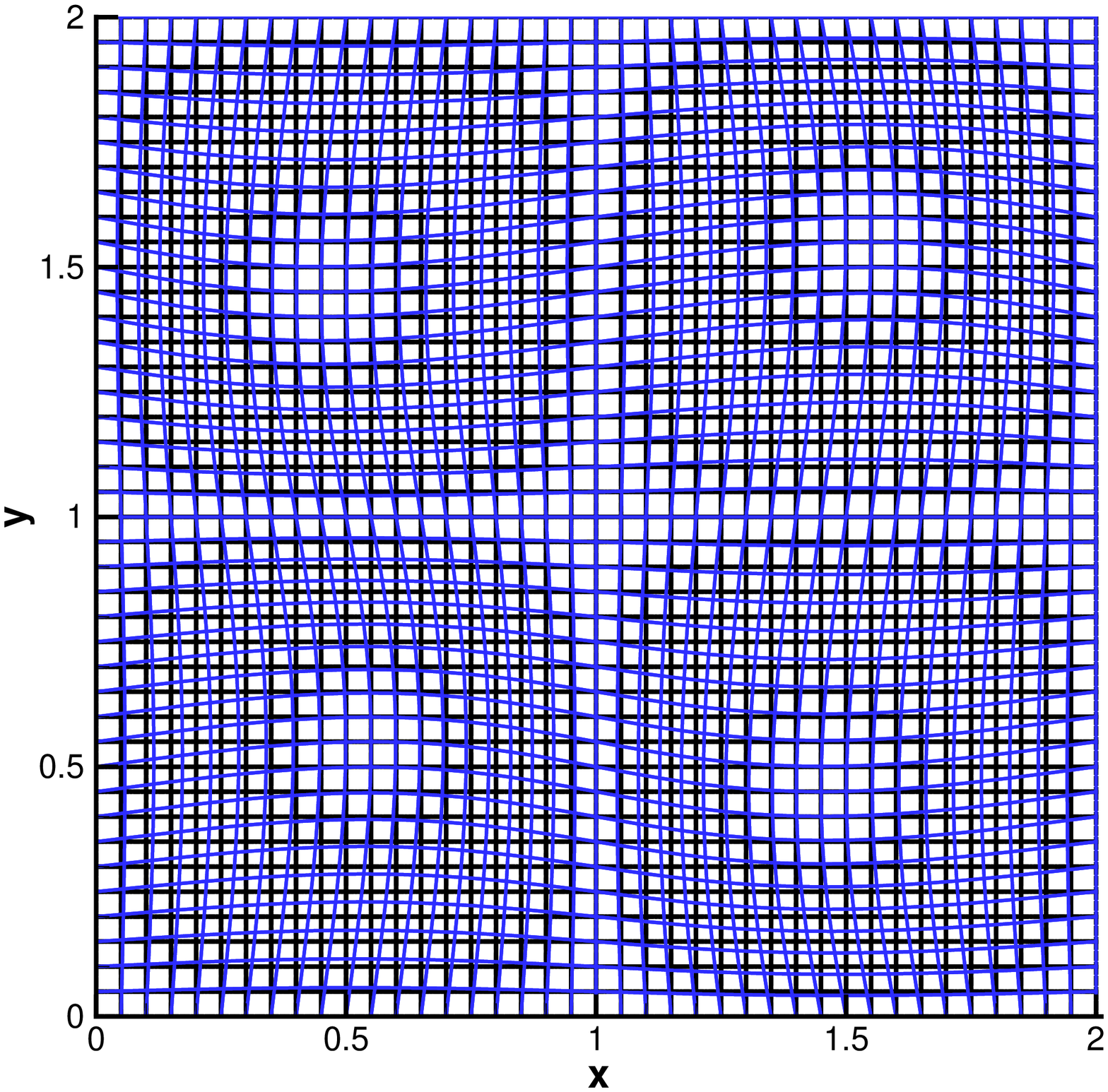}
\includegraphics[width=0.475\textwidth]{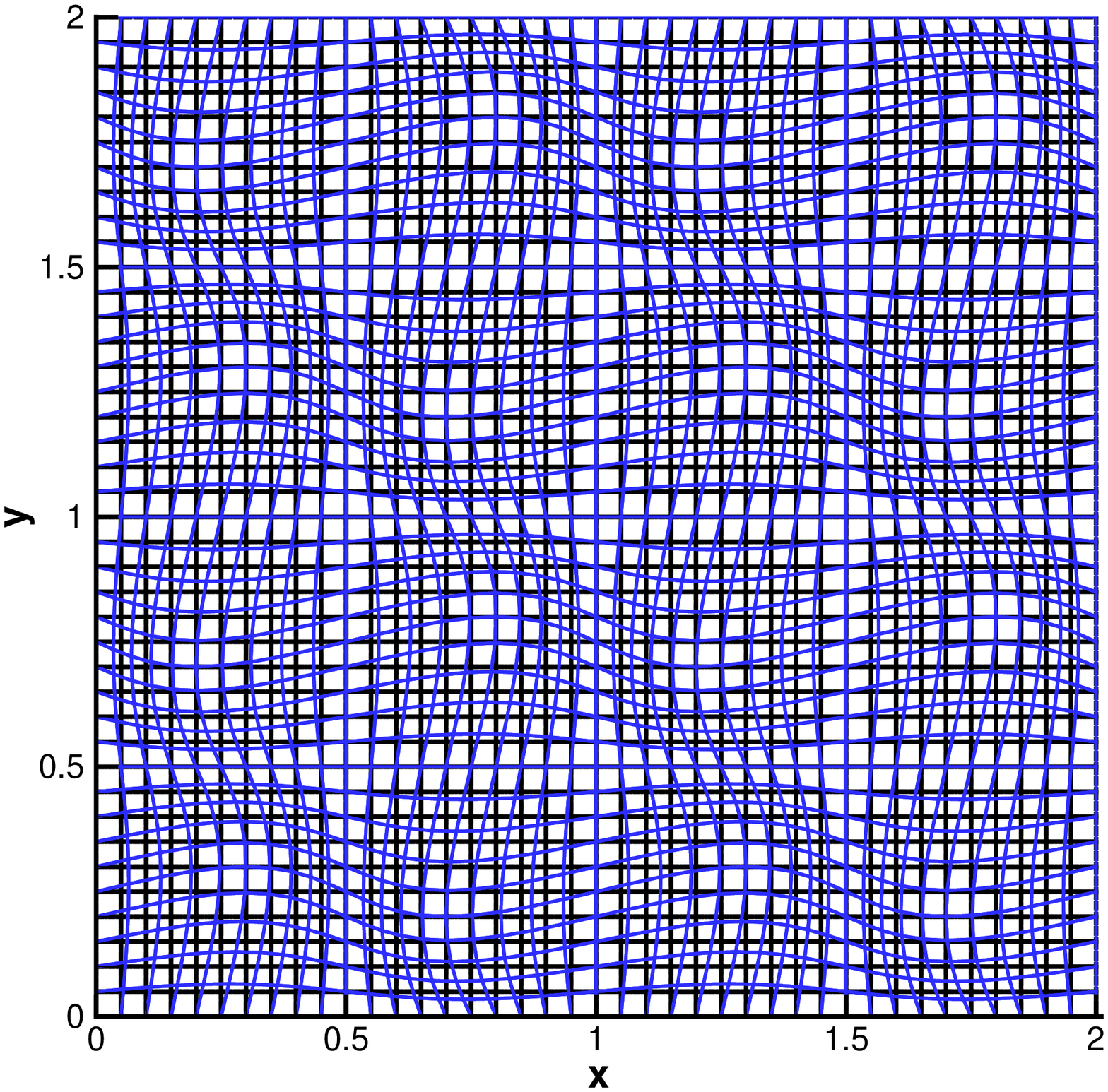}
\caption{\label{accuracy-mesh} Accuracy test: computational mesh
with $40^2$ cells at $t=0.5$ for advection of density perturbation
for moving-mesh Type-I (left) and Type-II (right).}
\end{figure}

\begin{table}[!h]
\begin{center}
\def\temptablewidth{0.75\textwidth}
{\rule{\temptablewidth}{1.0pt}}
\begin{tabular*}{\temptablewidth}{@{\extracolsep{\fill}}c|cc|cc}
mesh     & $L^1$ error  &   Order   &  $L^2$ error &  Order \\
\hline
$20^2$   &    3.2472E-02 &  ~       &  1.7989E-02  &  ~  \\
$40^2$   &    4.2331E-03 &  2.93  &  2.3494E-03  &  2.93 \\
$80^2$   &    5.3331E-04 &  2.98  &  2.9571E-04  &  2.99 \\
$160^2$  &    6.6761E-05 &  2.99  &  3.7016E-05  &  2.99 \\
\end{tabular*}
{\rule{\temptablewidth}{1.0pt}}
\end{center}
\vspace{-5mm}\caption{\label{accuracy-1} Accuracy test: 2D advection
of density perturbation for moving-mesh Type-I.}
\begin{center}
\def\temptablewidth{0.75\textwidth}
{\rule{\temptablewidth}{1.0pt}}
\begin{tabular*}{\temptablewidth}{@{\extracolsep{\fill}}c|cc|cc}
  mesh   & $L^1$ error  &     Order   &  $L^2$ error &  Order    \\
\hline
$20^2$   &   3.3873E-02 &   ~      &   1.8770E-02 &   ~      \\
$40^2$   &   4.5521E-03 &  2.89  &   2.5410E-03 &   2.88 \\
$80^2$   &   5.7874E-04 &  2.97  &   3.2386E-04 &   2.97 \\
$160^2$  &   7.2641E-05 &  2.99  &   4.0674E-05 &   2.99 \\
\end{tabular*}
{\rule{\temptablewidth}{1.0pt}}
\end{center}
\vspace{-5mm}\caption{\label{accuracy-2} Accuracy test: 2D advection
of density perturbation for moving-mesh Type-II.}
%\begin{center}
%\def\temptablewidth{0.75\textwidth}
%{\rule{\temptablewidth}{1.0pt}}
%\begin{tabular*}{\temptablewidth}{@{\extracolsep{\fill}}c|cc|cc}
%mesh     & $L^1$ error  &    Order    &  $L^2$ error &  Order   \\
%\hline
%$20^2$   &  3.1442E-02  &    ~    &  1.7398E-02   &   ~     \\
%$40^2$   &  4.0849E-03  &  2.9443 &  2.2654E-03   &  2.9892 \\
%$80^2$   &  5.1443E-04  &  2.9410 &  2.8496E-04   &  2.9909 \\
%$160^2$  &  6.4388E-05  &  2.9980 &  3.5669E-05   &  2.9980 \\
%\end{tabular*}
%{\rule{\temptablewidth}{1.0pt}}
%\end{center}
%\vspace{-5mm}\caption{\label{accuracy-3} Accuracy test: 2D advection
%of density perturbation with stationary meshes.}
\end{table}

For the accuracy with moving-mesh velocity, two types of time
dependent meshes are considered as follows
\begin{align*}
\text{Type-I}: \begin{cases}
\displaystyle x(t)=x_0+0.05\sin\pi t\sin\pi x_0\sin\pi y_0,\\
\displaystyle y(t)=y_0+0.05\sin\pi t\sin\pi x_0\sin\pi y_0.
\end{cases}
\end{align*}
and
\begin{align*}
\text{Type-II}: \begin{cases}
\displaystyle x(t)=x_0+0.05\sin\pi t\sin2\pi x_0\sin2\pi y_0,\\
\displaystyle y(t)=y_0+0.05\sin\pi t\sin2\pi x_0\sin2\pi y_0.
\end{cases}
\end{align*}
where $(x_0, y_0)$ is denoted as the initial mesh. The mesh velocity
can be directly given during a time step. Initially, $N\times N$
cells are distributed uniformly. The computational meshes with
largest deformation at $t=0.5$ are shown in Fig.\ref{accuracy-mesh}
as reference. The $L^1$ and $L^2$ errors and orders of accuracy at
$t=2$ with $N^2$ cells are presented in Tab.\ref{accuracy-1} and
Tab.\ref{accuracy-2} for two types of moving-mesh with linear
weights. The expected accuracy is well kept during the moving-mesh
procedure.

\begin{table}[!h]
\begin{center}
\def\temptablewidth{0.85\textwidth}
{\rule{\temptablewidth}{1.0pt}}
\begin{tabular*}{\temptablewidth}{@{\extracolsep{\fill}}c|cc|cc}
~      & mesh velocity I & ~&  mesh velocity II \\
 mesh  & $L^1$ error  &  $L^2$ error  & $L^1$ error  &  $L^2$ error \\
\hline
$10^2$   &    4.6940E-15 & 3.1077E-15  &  5.2846E-15 & 3.2855E-15 \\
$20^2$   &    1.6520E-14 & 1.1142E-14  &  1.6967E-14 & 1.2542E-14 \\
$40^2$   &    4.2199E-14 & 2.8577E-14  &  6.9709E-14 & 4.3066E-14 \\
$80^2$   &    1.0889E-13 & 7.0537E-14  &  3.2411E-13 & 2.0728E-13
\end{tabular*}
{\rule{\temptablewidth}{1.0pt}}
\end{center}
\vspace{-5mm} \caption{\label{accuracy-4} Accuracy test: geometric
conservation law for mesh velocity Type-I and Type-II.}
\end{table}

\subsection{Geometric conservation law}
The geometric conservation law (GCL) \cite{GCL-1,GCL-2} is also
tested by the time dependent meshes given above. The GCL is mainly
about the maintenance of a uniform flow passing through a
non-uniform moving mesh. The initial condition for the
two-dimensional case is
\begin{align*}
\rho_0(x,y)=1,~p_0(x,y)=1,~U_0(x,y)=1,~V_0(x,y)=1.
\end{align*}
The periodic boundary conditions are adopted as well. The $L^1$ and
$L^2$ errors at $t=0.1$  with $N^2$ cells are given in
Tab.\ref{accuracy-4}  for two types of moving-mesh. The results show
that the errors reduce to the machine zero. The geometric
conservation law is well preserved by the current scheme.

\begin{figure}[!htb]
\centering
\includegraphics[width=0.475\textwidth]{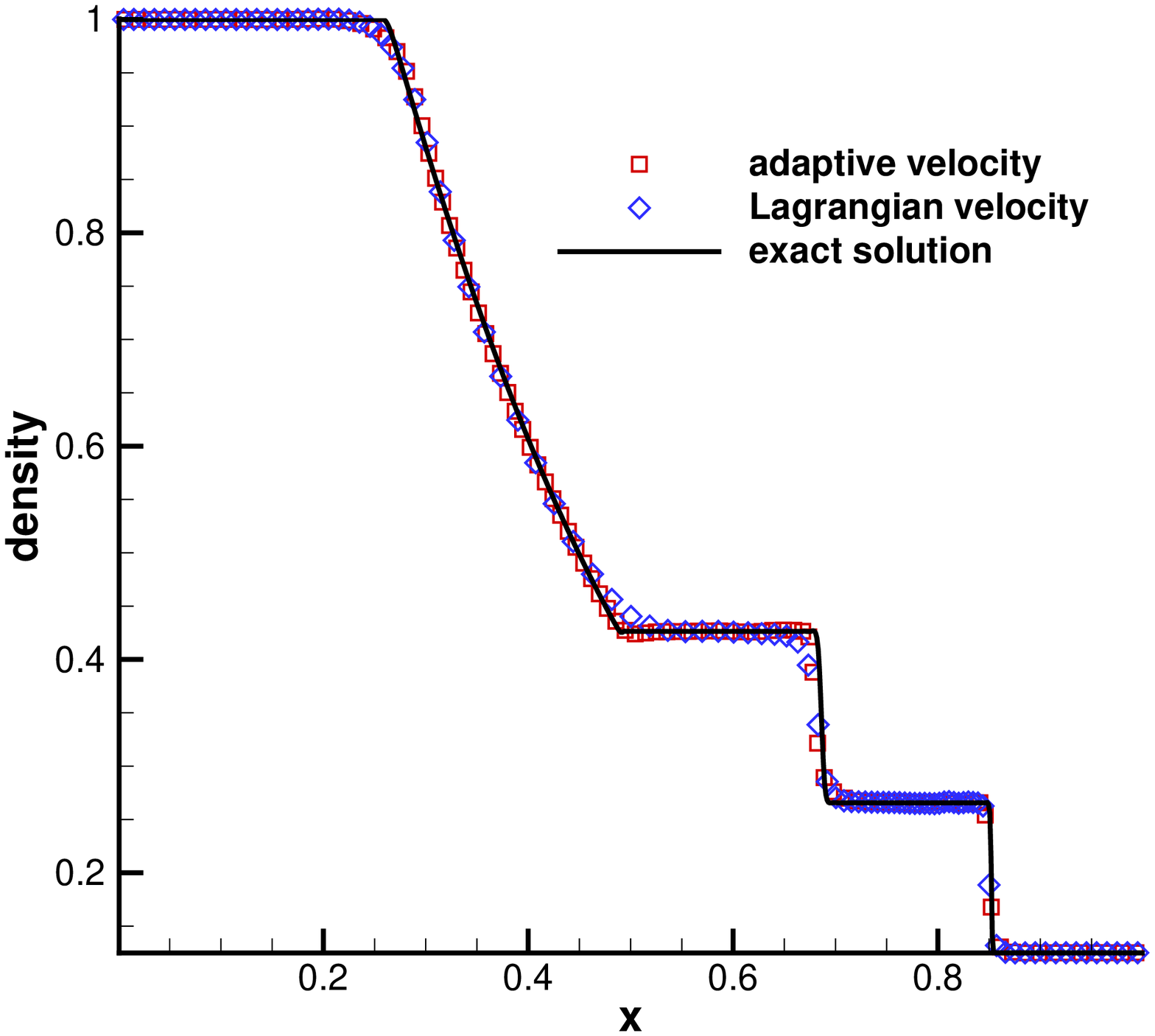}
\includegraphics[width=0.475\textwidth]{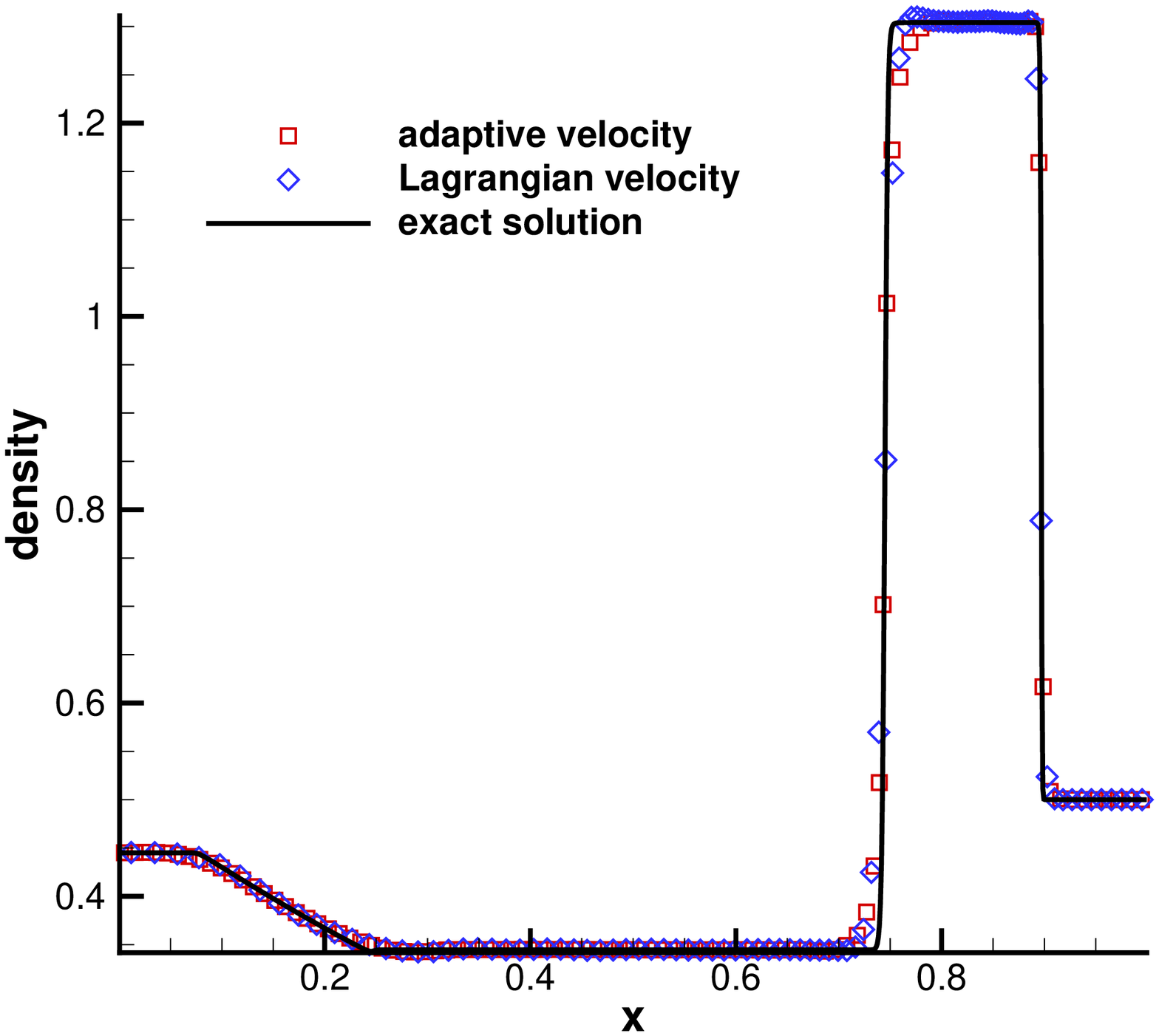}
\caption{\label{1d-riemann-1} One dimensional Riemann problem: the
density distributions for Sod problem at $t=0.2$ (left) and Lax
problem at $t=0.16$ (right).}
\end{figure}

\begin{figure}[!htb]
\centering
\includegraphics[width=0.475\textwidth]{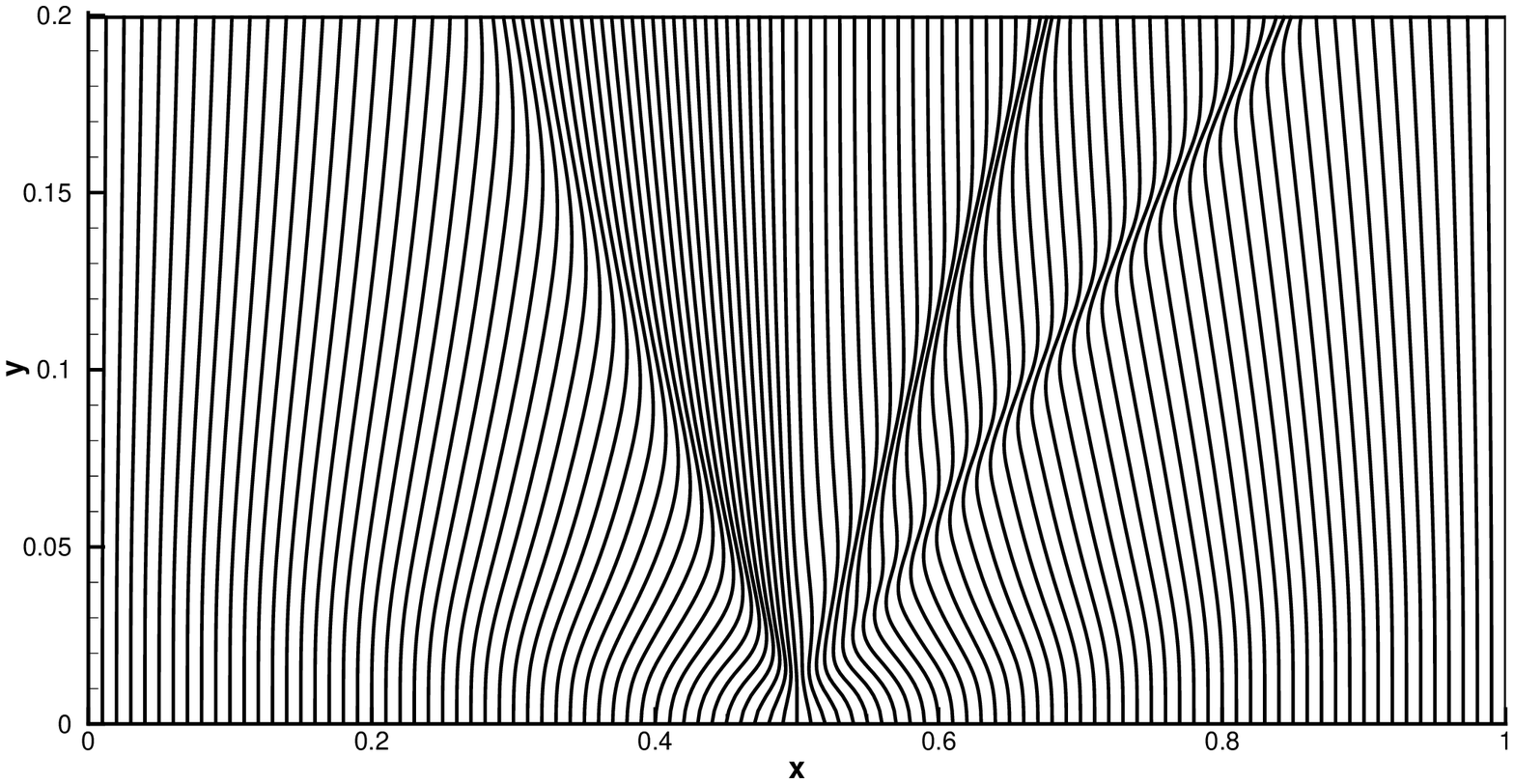}
\includegraphics[width=0.475\textwidth]{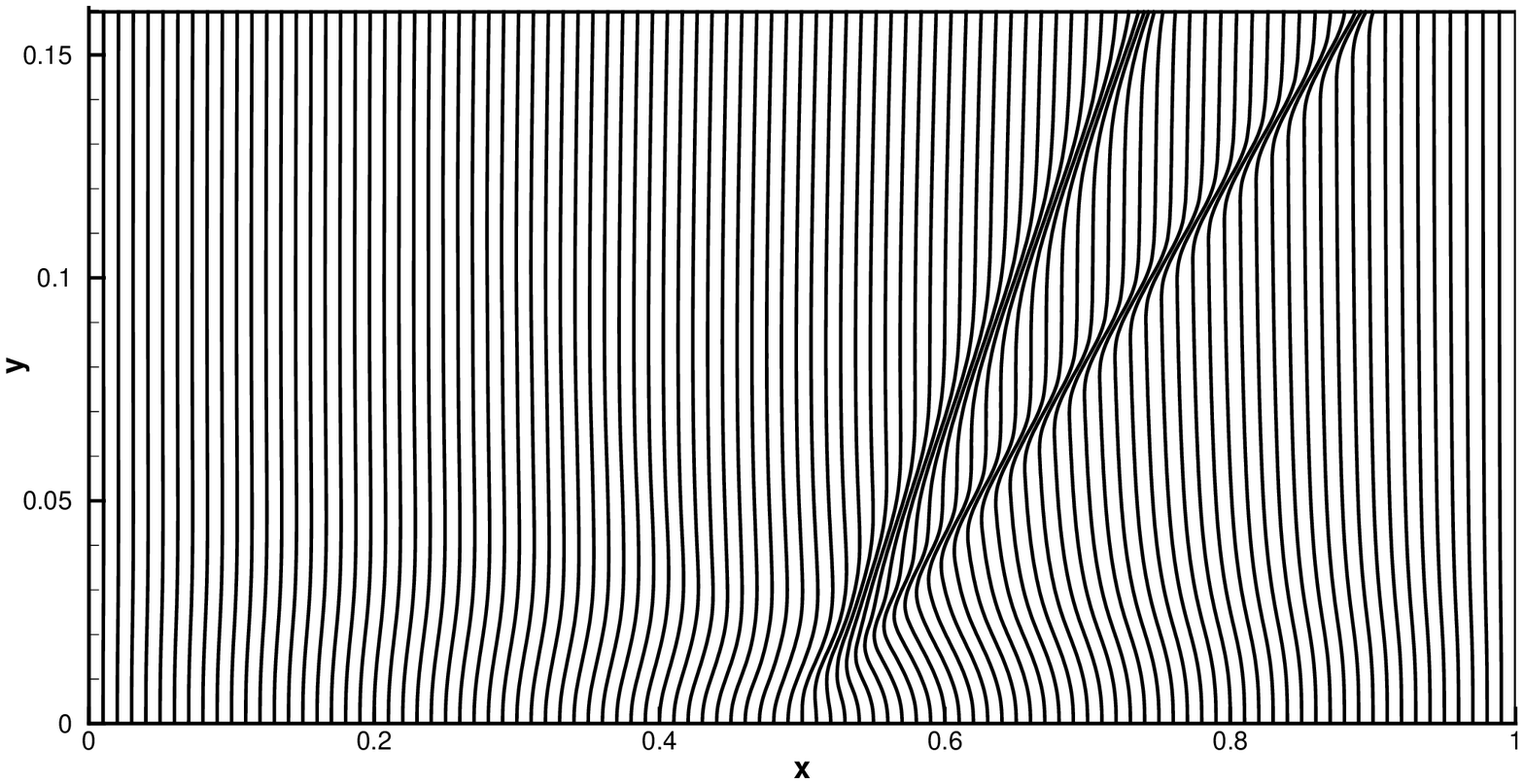}
\caption{\label{1d-riemann-2} One dimensional Riemann problem: the
mesh with adaptation velocity for Sod problem (left) and Lax problem
(right).}
\includegraphics[width=0.475\textwidth]{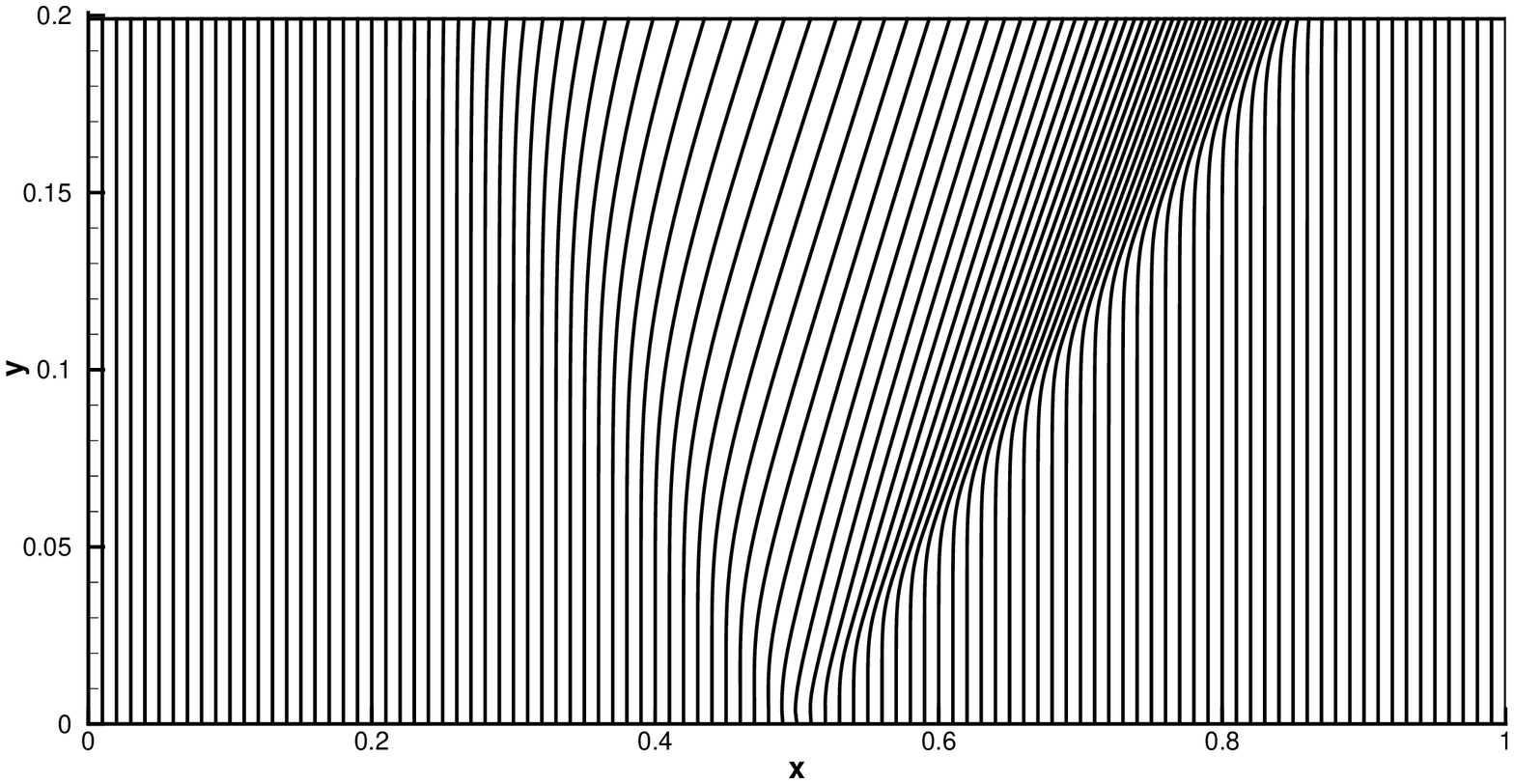}
\includegraphics[width=0.475\textwidth]{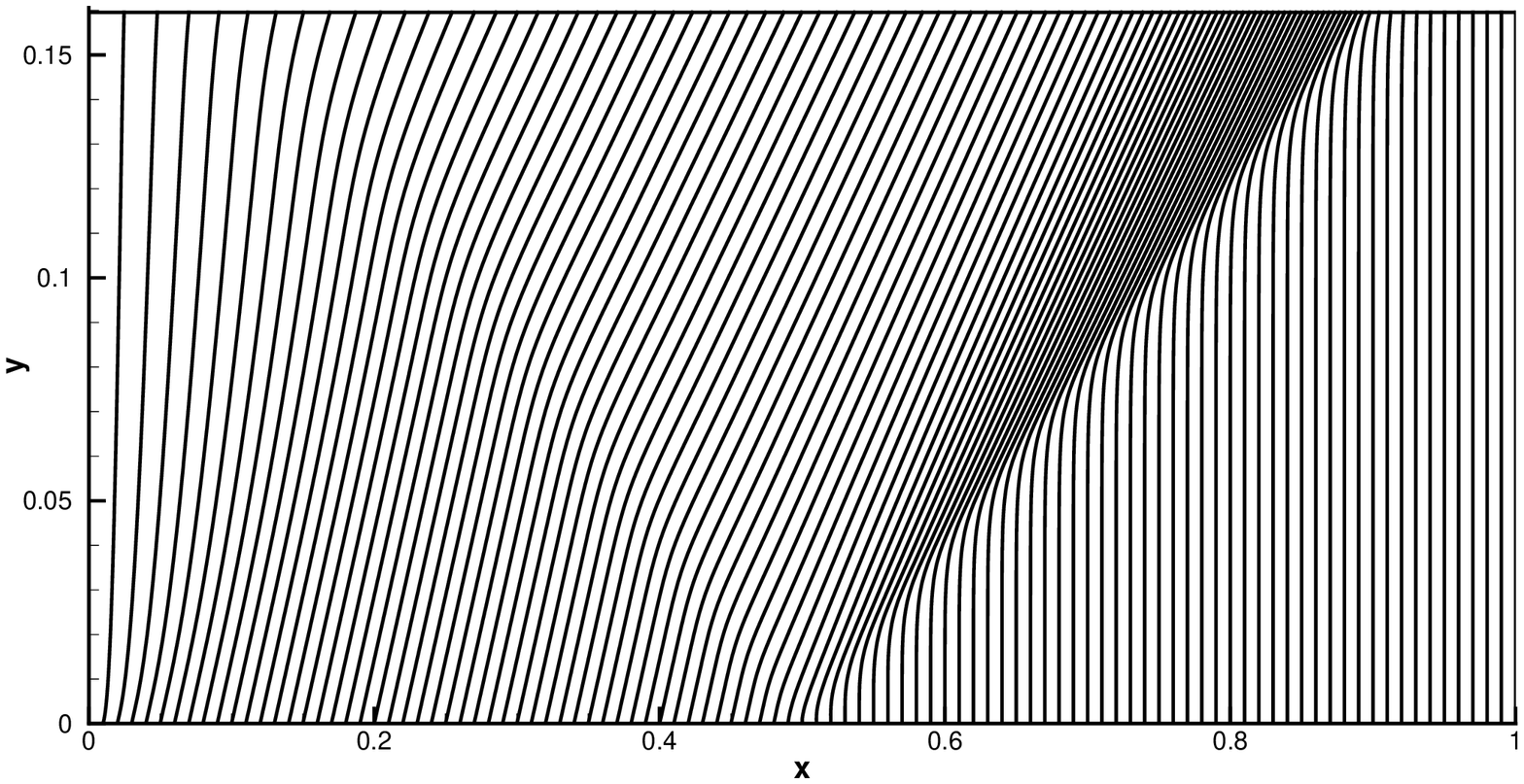}
\caption{\label{1d-riemann-3} One dimensional Riemann problem: the
mesh with Lagrangian velocity for Sod problem (left) and Lax problem
(right).}
\end{figure}

\subsection{One dimensional Riemann problems}
In this case, one-dimensional Riemann problems are tested by the
current scheme. The first one is the Sod problem, and the initial
condition is given as follows
\begin{equation*}
(\rho,U,p)=
\begin{cases}
(1,0,1), &0\leq x<0.5,\\
(0.125, 0, 0.1), &0.5\leq x\leq1.
\end{cases}
\end{equation*}
The second one is the Lax problem, and the initial condition is
given as follows
\begin{align*}
(\rho,U,p)=
\begin{cases}
(0.445,1.198, 3.528), &0\leq x<0.5\\
(0.5,0.5,0.571),  &0.5\leq x\leq1.
\end{cases}
\end{align*}
In the computation, these two cases are tested in the domain
$[0,1]\times[0,0.1]$ and non-reflection boundary condition is
adopted at the boundaries of computational domain. Initially,
$100\times10$ cells are equally distributed. The adaptation velocity
and the Lagrangian velocity are chosen as the mesh velocity. For the
adaptive procedure, the parameter $\alpha$ in the monitor function
takes $0.1$ for the Sod problem and $0.02$ for the Lax problem. The
density distributions for Sod problem at $t=0.2$ and Lax problem at
$t=0.16$ with $x=0$ are presented in Fig.\ref{1d-riemann-1}. The
history of mesh distribution  from adaptation velocity and Lagrangian
velocity are presented in Fig.\ref{1d-riemann-2}
and Fig.\ref{1d-riemann-3}. The numerical solutions agrees well with the exact
solution. Due the local mesh adaptation, the discontinuities are
well resolved by the current scheme.

\begin{figure}[!htb]
\centering
\includegraphics[width=0.7\textwidth]{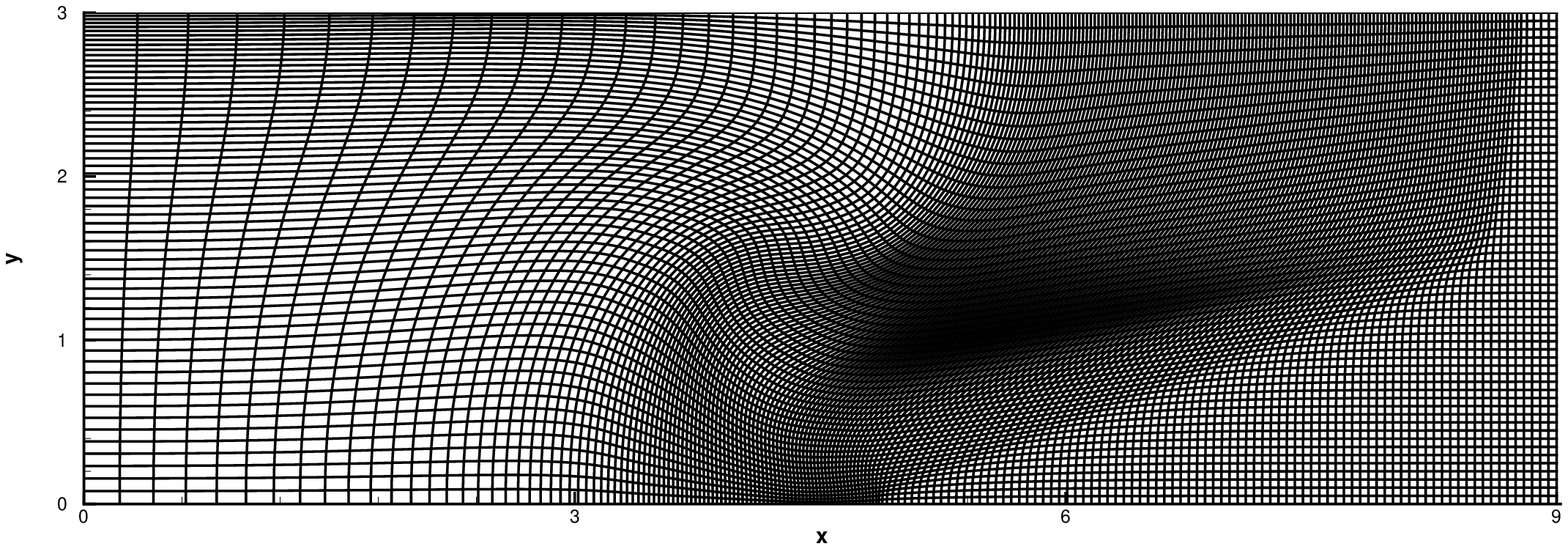}
\includegraphics[width=0.7\textwidth]{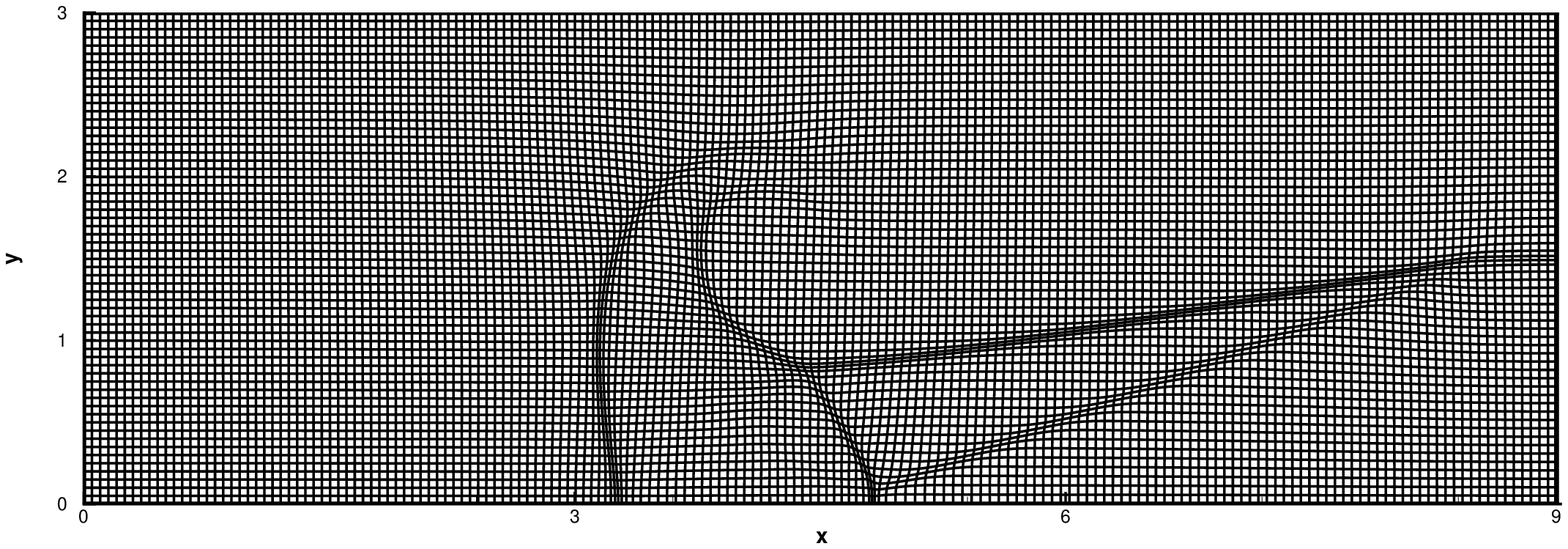}
\caption{\label{Triple-1} Triple-point problem: the mesh
distribution with $180\times60$ cells with Larangian velocity (top)
and adaptation velocity (bottom).}  \centering
\includegraphics[width=0.475\textwidth]{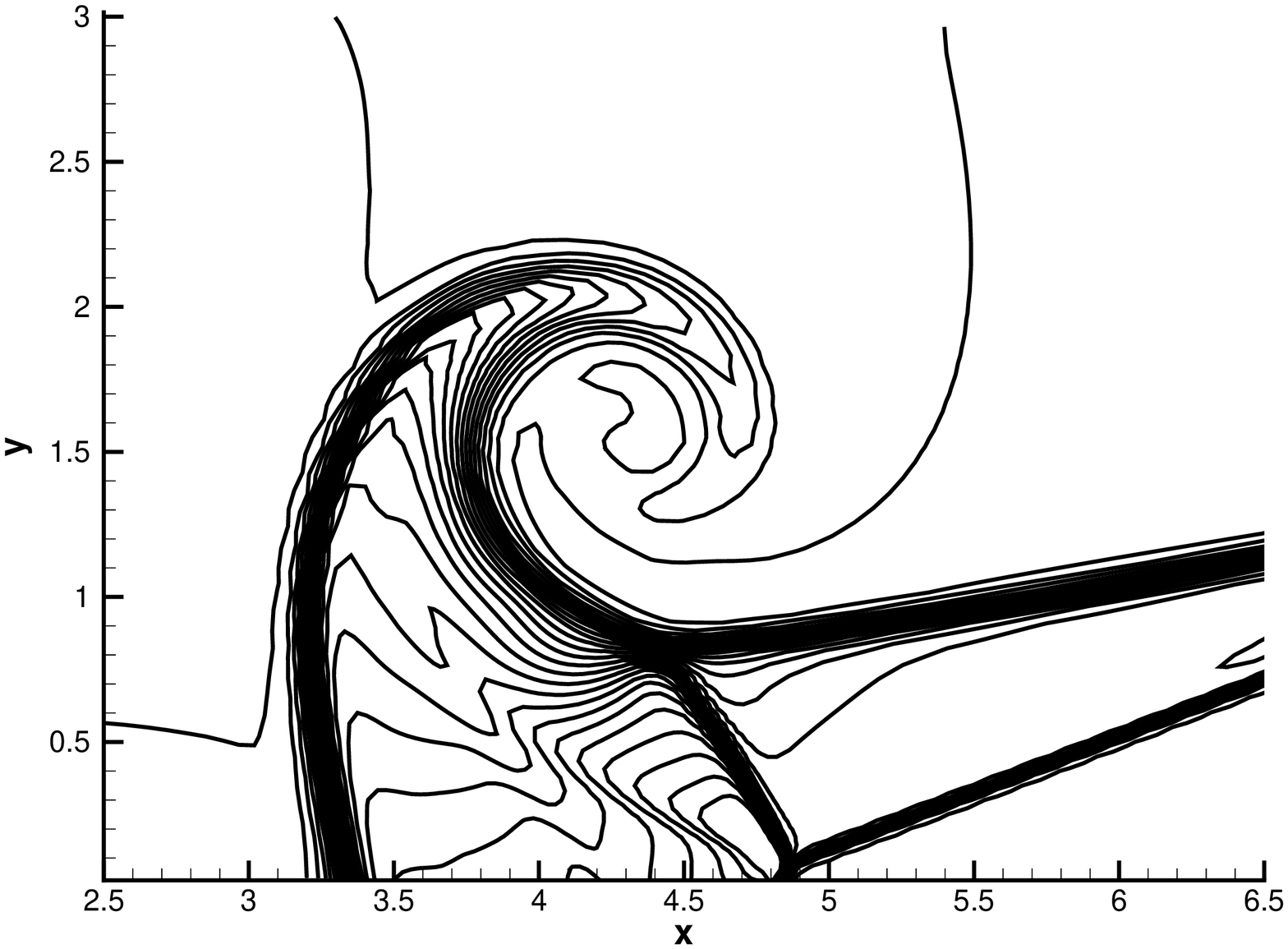}
\includegraphics[width=0.475\textwidth]{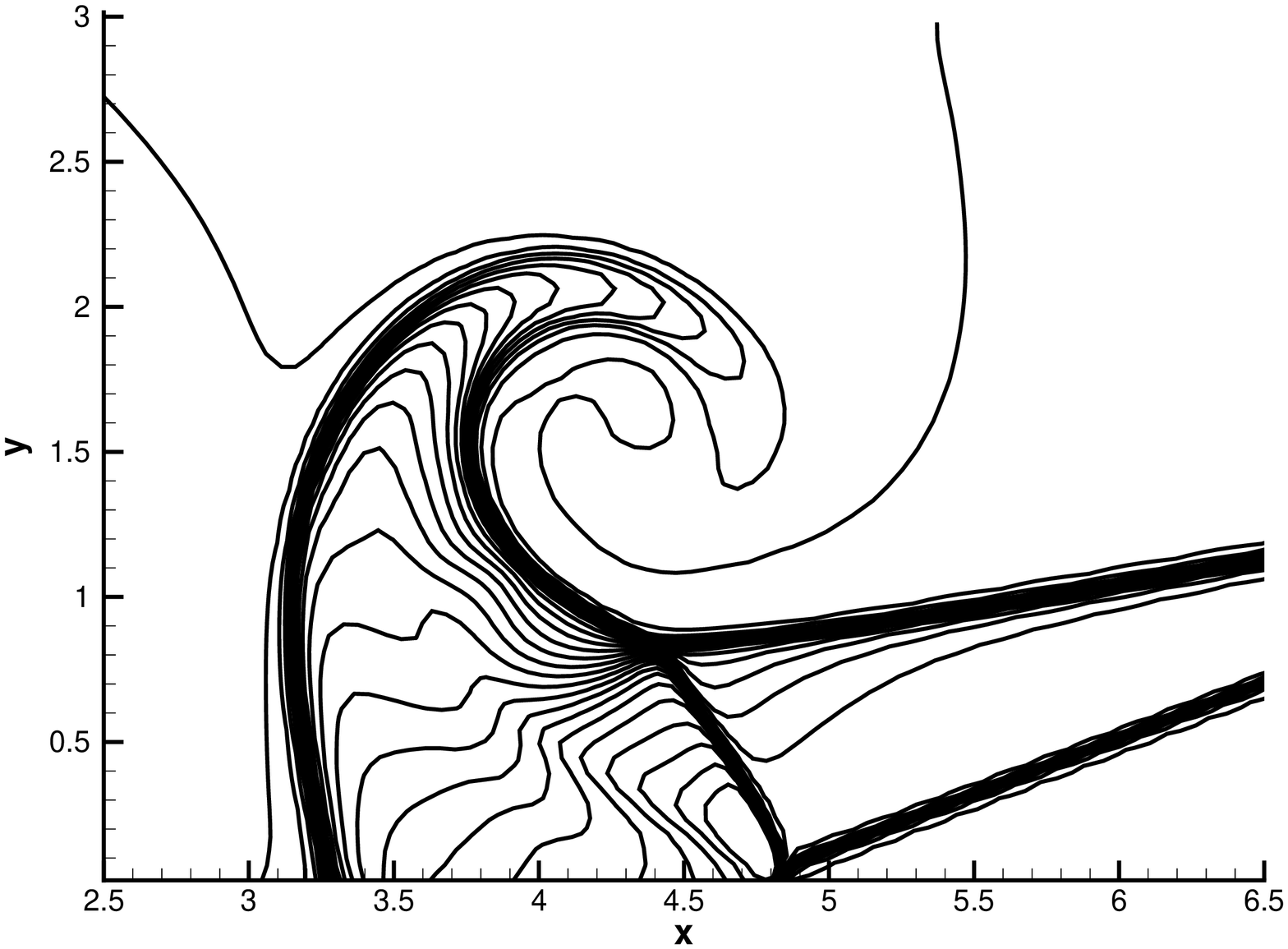}
\caption{\label{Triple-2} Triple-point problem: the density
distribution with Larangian velocity (left) and (right) adaptation
velocity with $180\times60$ cells.}
\end{figure}

\subsubsection{Triple-point problem}
The triple-point problem was widely used to validate to the
performance of Larangian and ALE methods with large mesh deformation
\cite{triple-point1, triple-point2}. The initial condition is given
as follows
\begin{align*}
(\rho,U,V,p)=\begin{cases}
(1,0,0,1), &(x,y)\in D_1=[0,1]\times[0,3],\\
(0.125,0,0,0.1), &(x,y)\in D_2=[1,9]\times[0,1.5],\\
(1,0,0,0.1), &(x,y)\in D_3=[1,9]\times[1.5,3].
\end{cases}
\end{align*}
The non-reflective boundary conditions are used at all boundaries.
Left from $x=1$, a high pressure is located in $D_1$, which
generates a shock wave propagating to the right, and no waves are
generated at the beginning at the domain $D_2\cup
D_3$. Initially, $3N\times N$ cells are equally distributed. The
adaptation velocity and Lagrangian velocity are chosen as the mesh
velocity. For the adaptive procedure, the parameter $\alpha$ in the
monitor function takes $0.05$. The mesh distribution with
$180\times60$ cells for Larangian and adaptation velocity at $t=4$
are presented in Fig.\ref{Triple-1}. The density distribution with
$180\times60$ cells for Larangian and adaptation velocity are given
in Fig.\ref{Triple-2}, where a vortex appears and swirls the flows
around the triple point due to the different shock speeds over the
vertical line.

\begin{figure}[!htb]
\centering
\includegraphics[width=0.45\textwidth]{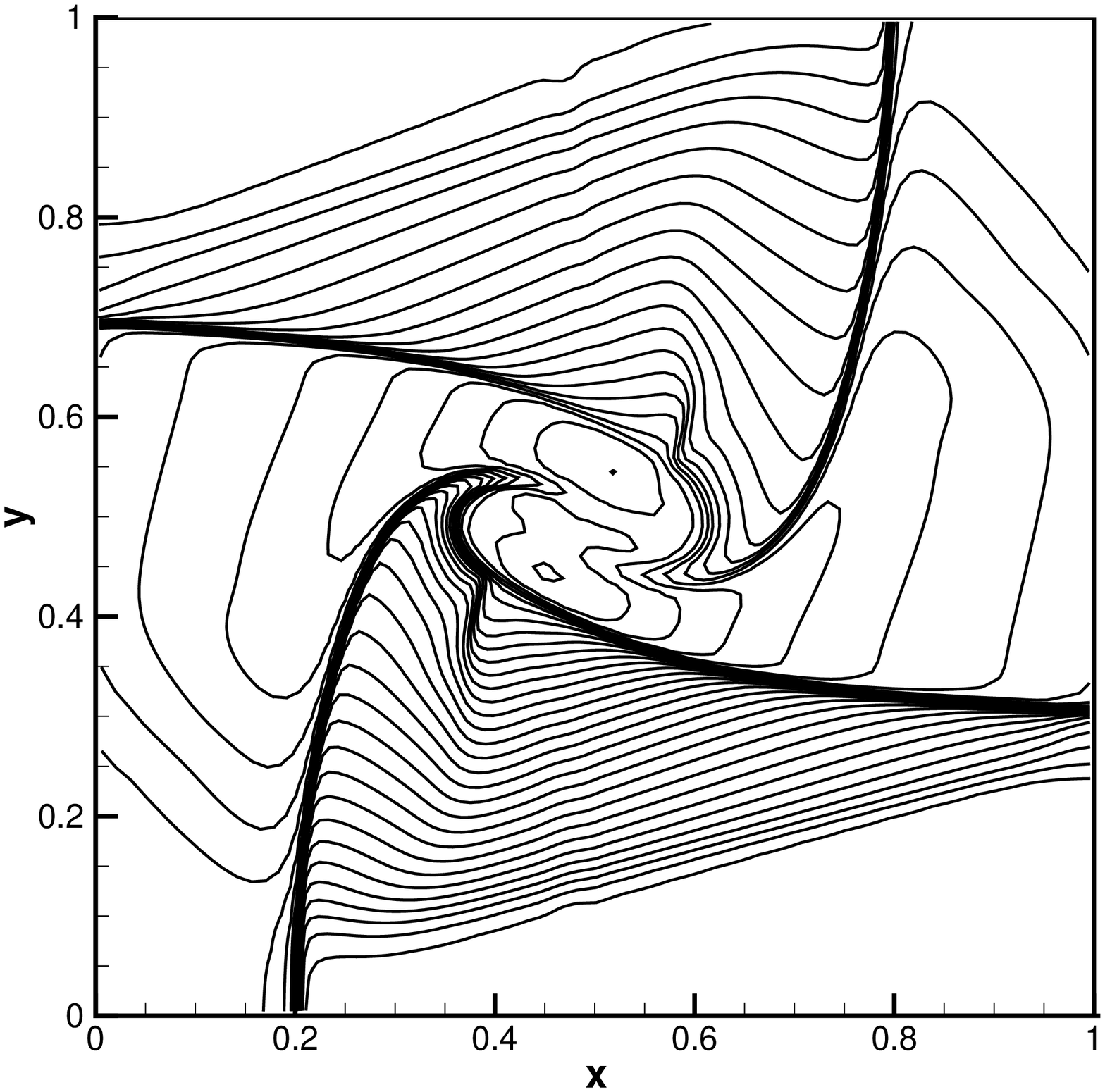}
\includegraphics[width=0.45\textwidth]{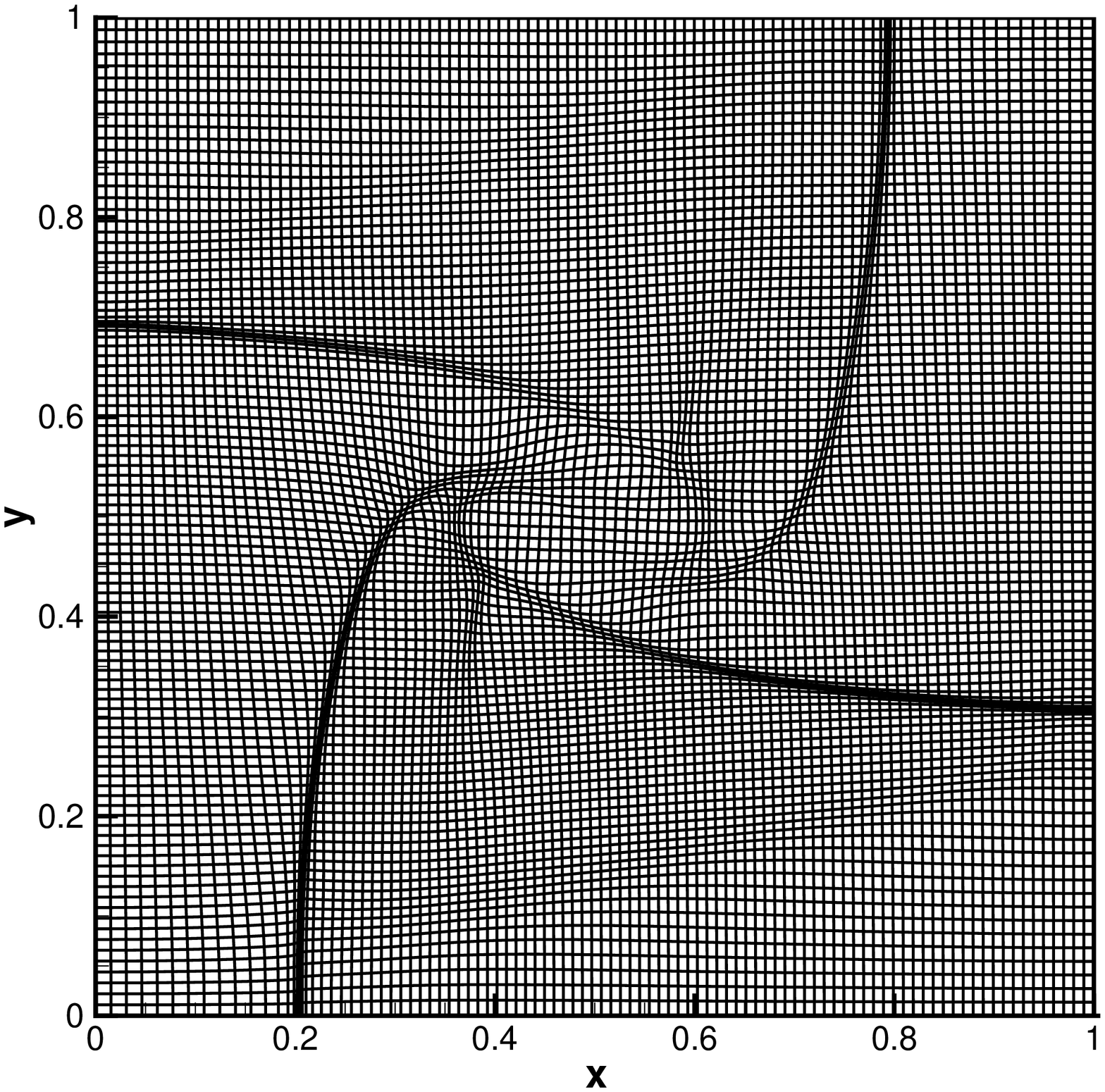}
\caption{\label{2d-riemann-a1} Two dimensional Riemann problem: the
density distribution (left) and mesh distribution (right) for four
contact discontinuity interaction with $100\times100$ cells.}
\centering
\includegraphics[width=0.45\textwidth]{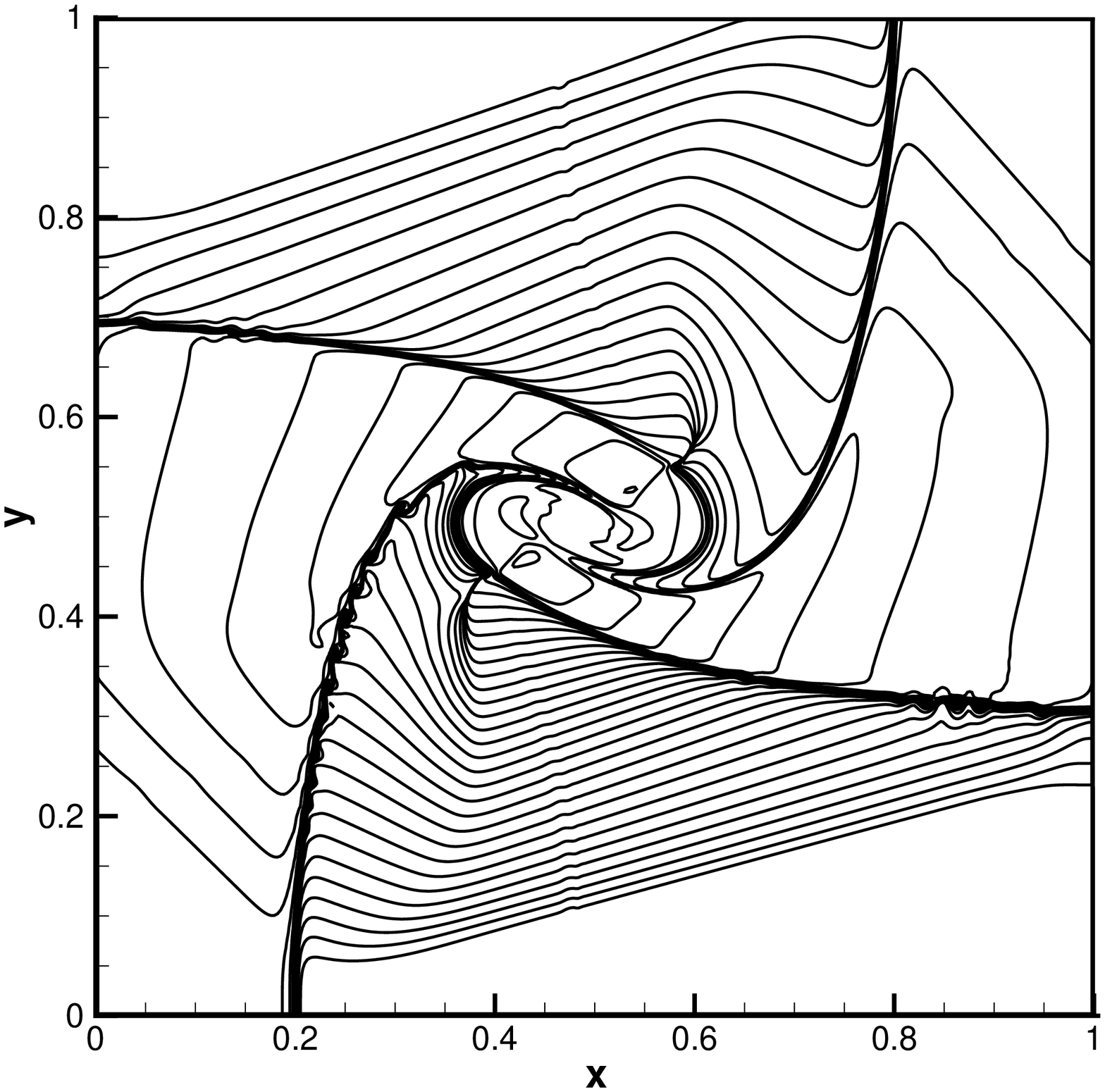}
\includegraphics[width=0.45\textwidth]{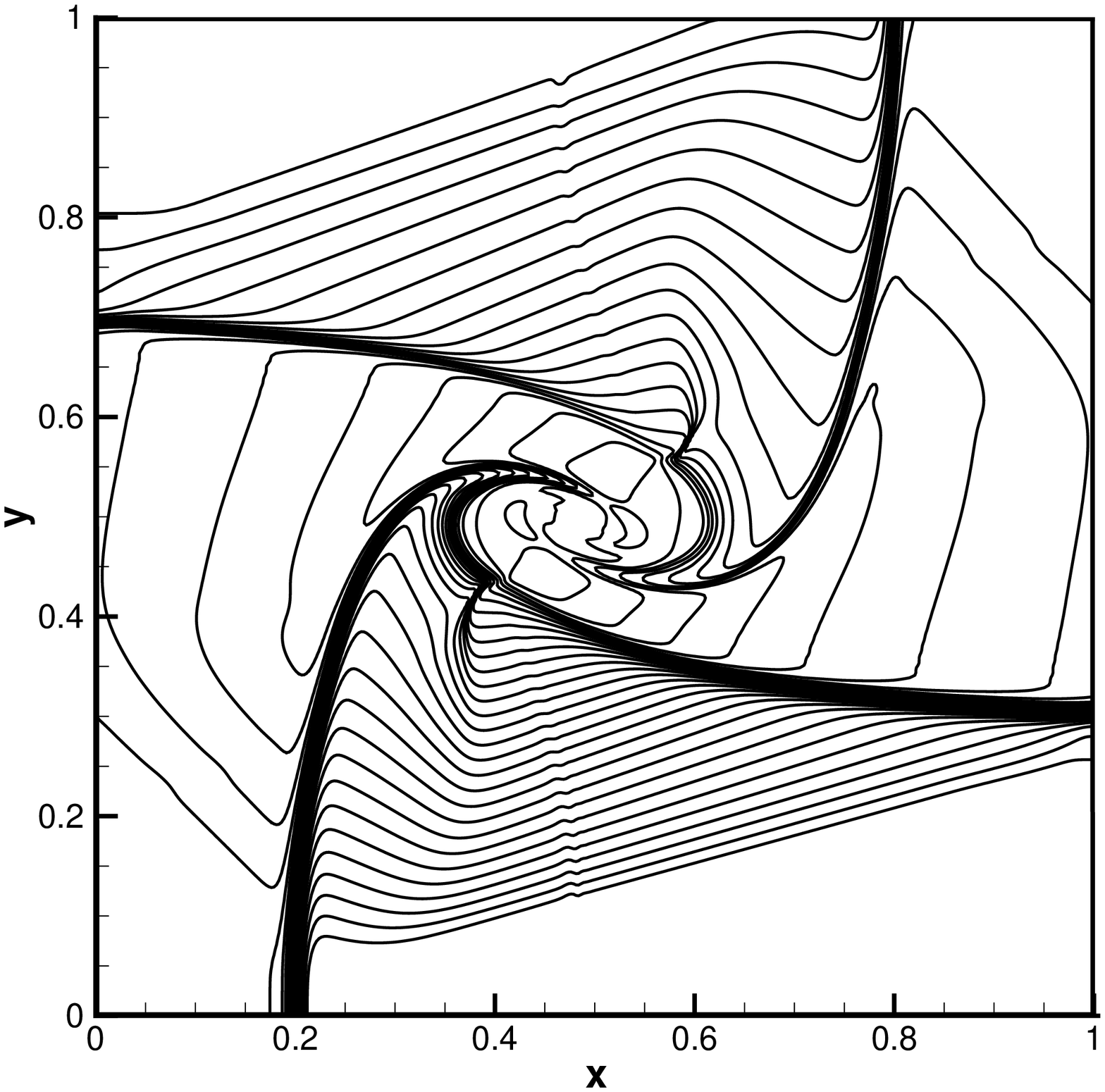}
\caption{\label{2d-riemann-a2} Two dimensional Riemann problem: the
density distribution on moving-mesh (left) and stationary mesh
(right) for four contact discontinuity interaction with
$400\times400$ cells.}
\end{figure}

\subsection{Two-dimensional Riemann problems}
In this case, the two-dimensional Riemann problems for Euler
equations are presented \cite{Case-Lax}. The computation domain is
$[0,1]\times[0,1]$ and the non-reflection condition is used for the
boundaries, and $N\times N$ cells are equally distributed initially.
The adaptation velocity are chosen as the mesh velocity. The initial
condition is given as follows
\begin{align*}
(\rho,U,V,p)&= \left\{\begin{aligned}
&(1 ,0.75,-0.5,1), &x>0.5,y>0.5,\\
&(2,0.75,0.5,1), &x<0.5,y>0.5,\\
&(1,-0.75,0.5,1), &x<0.5,y<0.5,\\
&(3,-0.75,-0.5,1), &x>0.5,y<0.5.
\end{aligned}
\right.
\end{align*}
This case is the interaction of planar contact discontinuity for
vortex sheets with same signs $J_{12}^- J_{32}^- J_{41}^- J_{34}^-$,
where $J^-_{lr}$ represents the negative contact discontinuity
connecting the $l$ and $r$ areas
\begin{align*}
J_{lr}^-: w_l=w_r, p_l=p_r, w_l'\geq w_r',
\end{align*}
where $w_l, w_r$ are the normal velocity and $w_l', w_r'$ are the
tangential velocity. The parameter $\alpha$ in the monitor function
takes $0.1$. The density distribution and computational mesh with
$100\times100$ cells are given in Fig.\ref{2d-riemann-a1}. The
density distribution with $400\times400$ cells is given in
Fig.\ref{2d-riemann-a2}. The instantaneous interaction of contacts
results in a complex wave pattern, and the instability appears due
to the local mesh adaptation. As a reference, the numerical results
with stationary mesh is given in Fig.\ref{2d-riemann-a2} as well.
The nonlinear combination of linear polynomial smears the
discontinuities.

\begin{figure}[!h]
\centering
\includegraphics[width=0.475\textwidth]{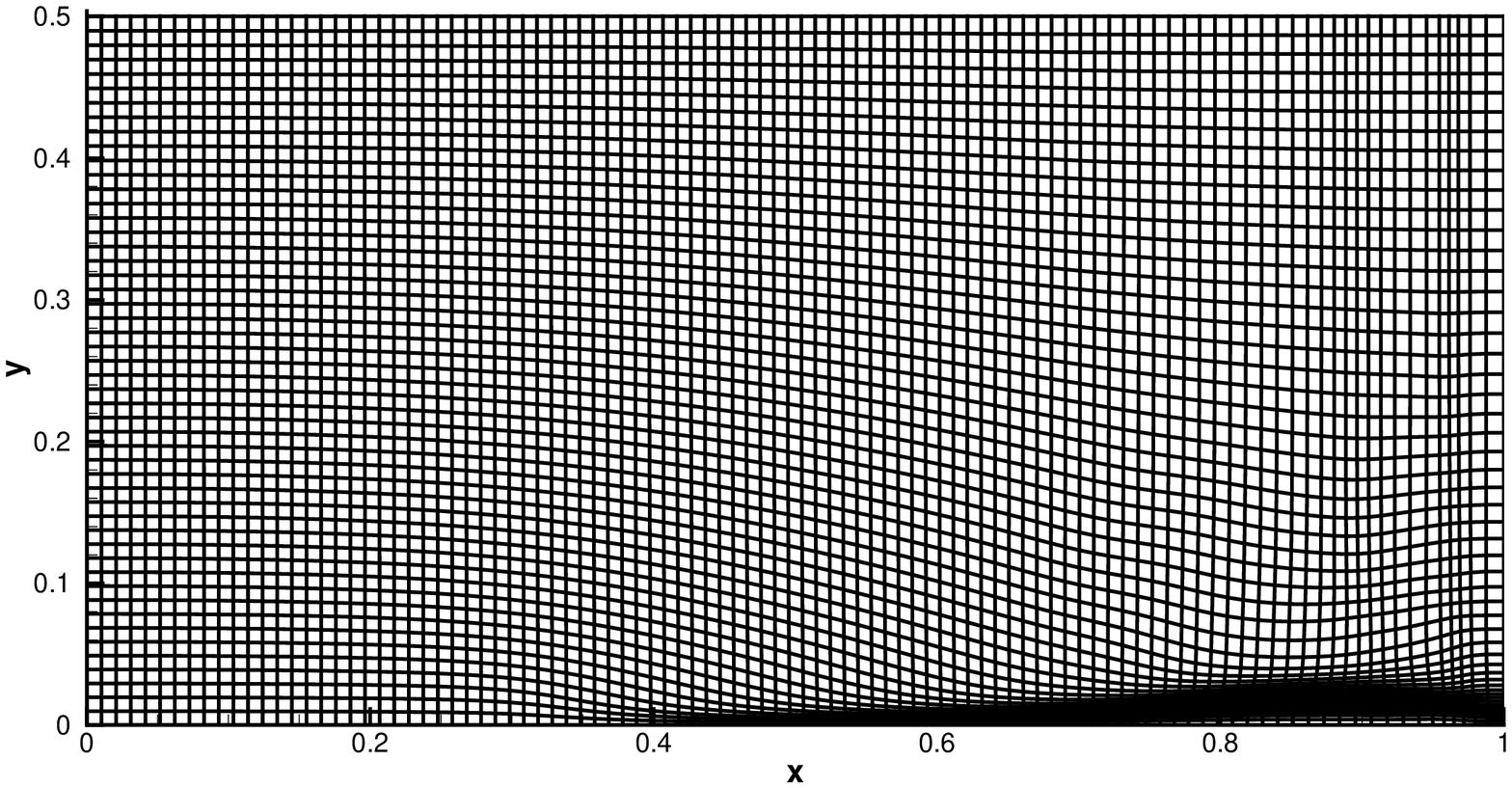}
\includegraphics[width=0.475\textwidth]{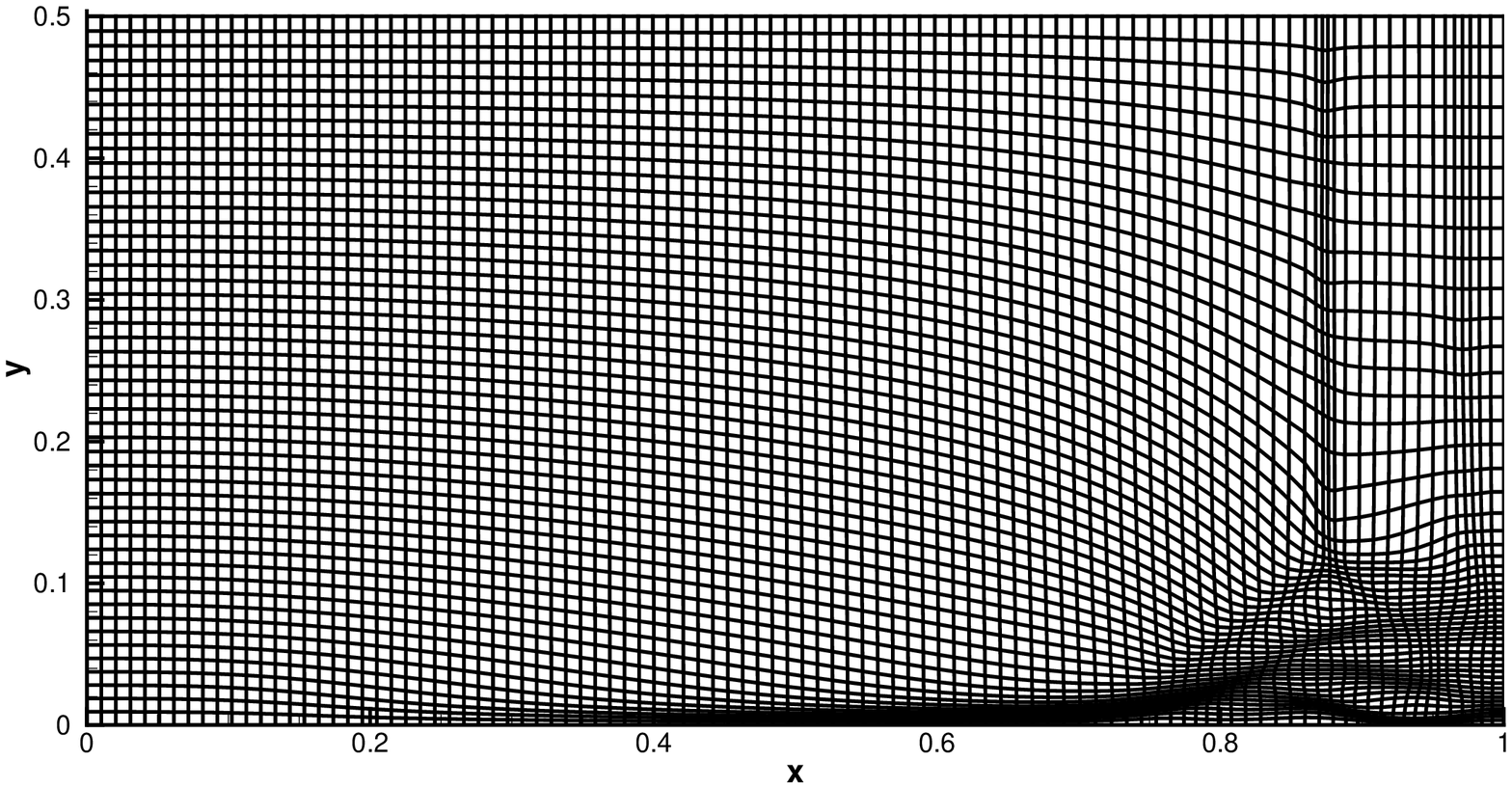}\\
\includegraphics[width=0.475\textwidth]{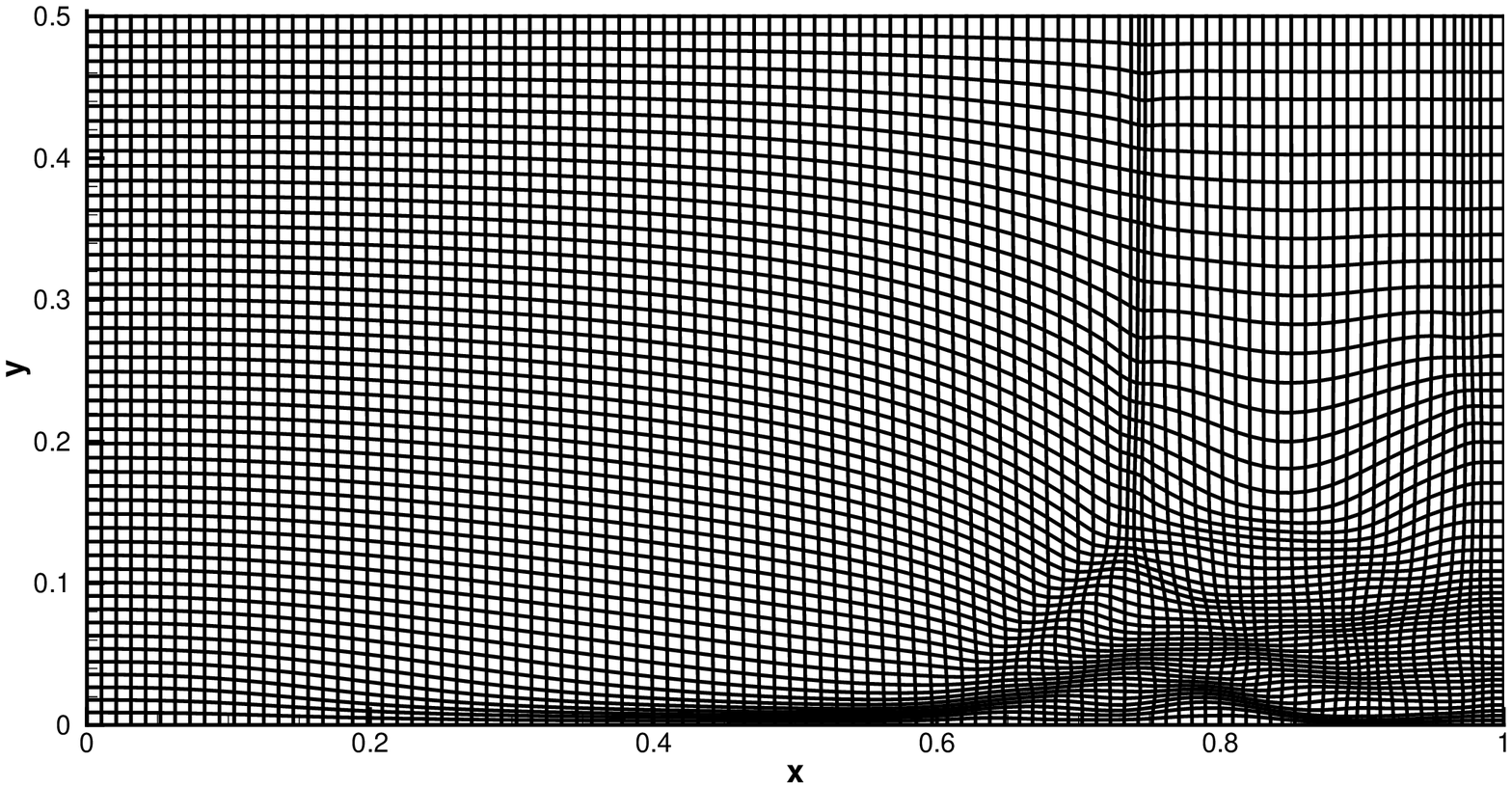}
\includegraphics[width=0.475\textwidth]{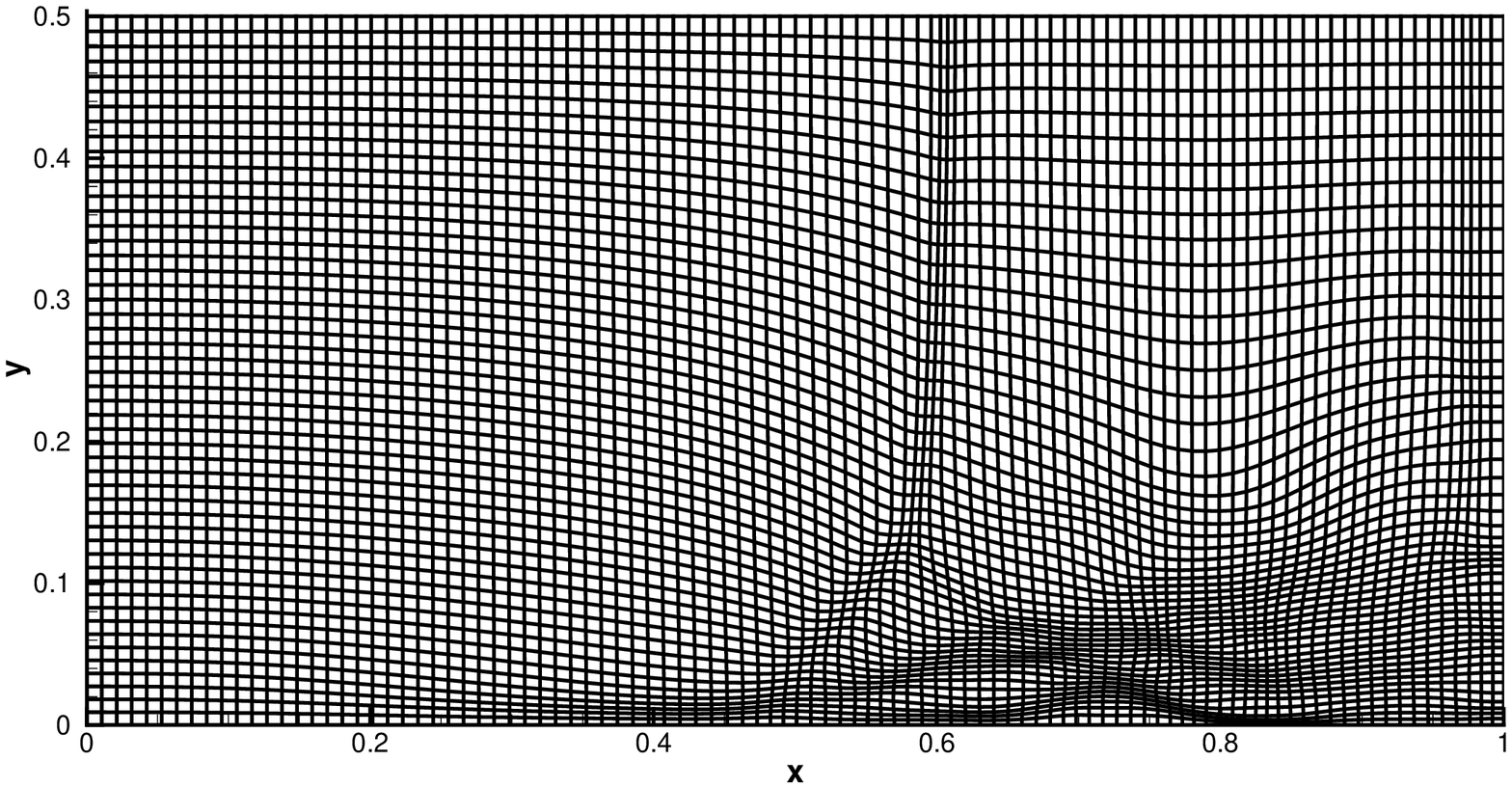}
\caption{\label{viscous-shock-1} Viscous shock tube: the
computational mesh at $t=0.25, 0.5, 0.75$ and $1$.}
\end{figure}

\subsection{Viscous shock tube}
This problem was introduced to test the performances of current
scheme for viscous flows \cite{Case-Daru}. In this case, an ideal
gas is at rest in a two-dimensional unit box $[0,1]\times[0,1]$. A
membrane located at $x=0.5$ separates two different states of the
gas and the dimensionless initial states are
\begin{equation*}
(\rho,U,p)=\left\{\begin{aligned}
&(120, 0, 120/\gamma), \ \ \ &  0<x<0.5,\\
&(1.2, 0, 1.2/\gamma),  & 0.5<x<1,
\end{aligned} \right.
\end{equation*}
where $\gamma=1.4$, Reynolds number $\mbox{Re} =200$ and Prandtl
number $\mbox{Pr} =0.73$. In the computation, this case is tested in
the physical domain $[0, 1]\times[0, 0.5]$, a symmetric boundary
condition is used on the top boundary $x\in[0, 1], y=0.5$.  Non-slip
boundary condition for velocity, and adiabatic condition for
temperature are imposed at solid wall boundaries. The membrane is
removed at time zero and wave interaction occurs. A boundary layer
is generated at the beginning, and a shock moves to the right
followed by a contact discontinuity. After reflected by the end
wall, the discontinuities interact with the boundary layer.

\begin{table}[!h]
\begin{center}
\def\temptablewidth{1.0\textwidth}
{\rule{\temptablewidth}{0.5pt}}
\begin{tabular*}{\temptablewidth}{@{\extracolsep{\fill}}ccccc}
Scheme & AUSMPW+ \cite{Case-Kim} & M-AUSMPW+ \cite{Case-Kim} & WENO-GKS & GKS-ALE  \\
\hline
Height  & 0.163   &  0.168    & 0.171  &  0.164    \\
\end{tabular*}
{\rule{\temptablewidth}{0.5pt}}
\end{center}
\vspace{-4mm} \caption{\label{height}  Viscous shock tube:
comparison of primary vortex heights among different schemes.}
\end{table}

\begin{figure}[!h]
\centering
\includegraphics[width=0.65\textwidth]{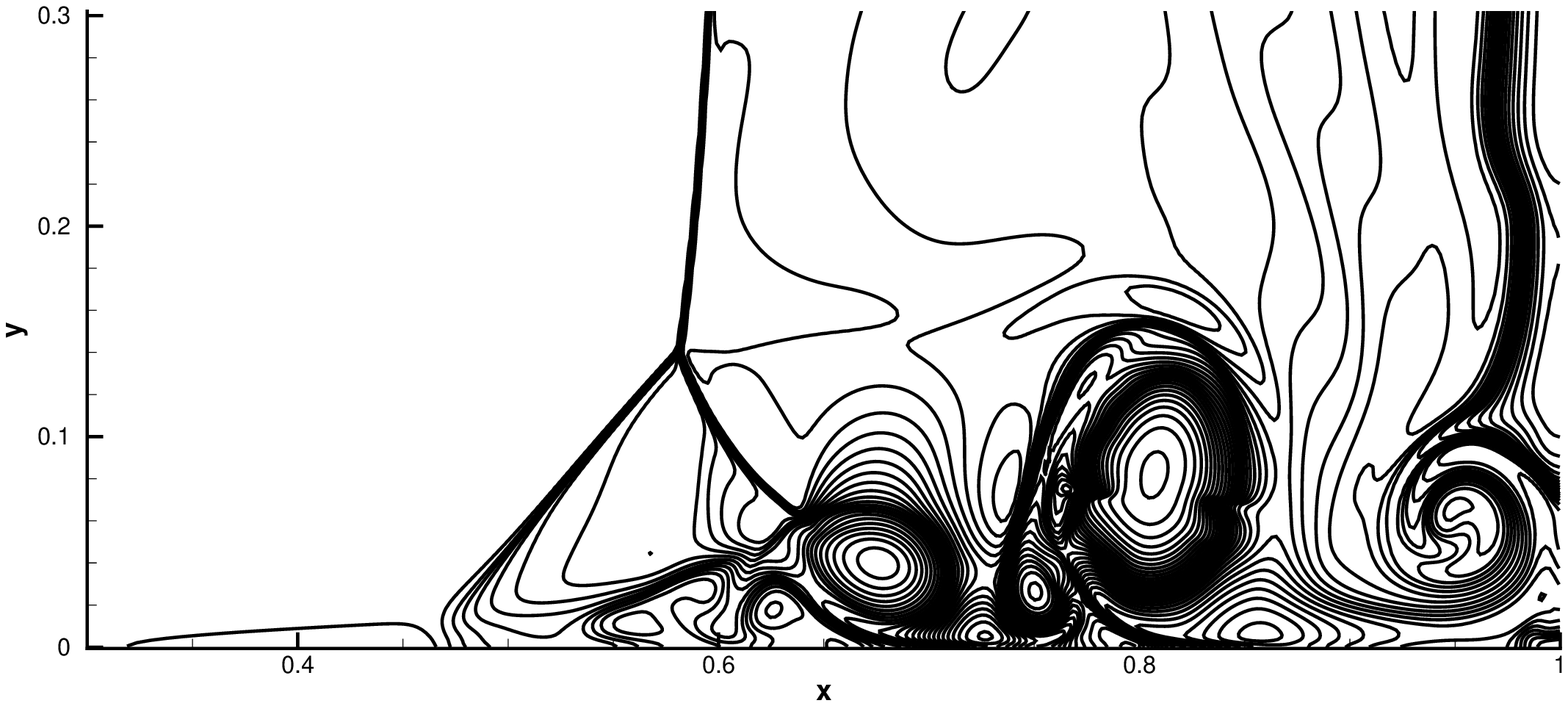}
\includegraphics[width=0.65\textwidth]{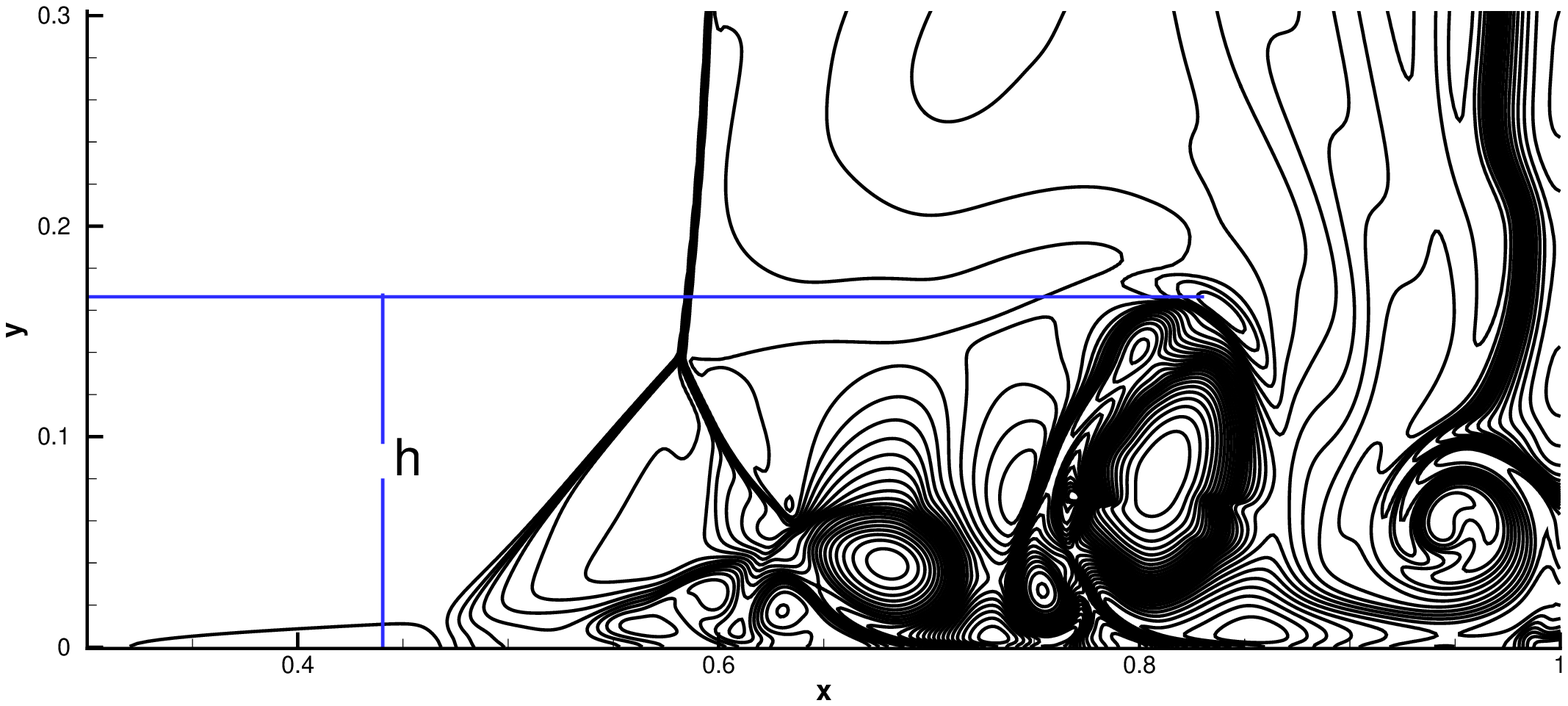}
\caption{\label{viscous-shock-2} Viscous shock tube: the density
distributions on the mesh with $500\times250$ and $700\times350$
cells.}
\end{figure}

To well resolve the complicated two-dimensional
shock/shear/boundary-layer interactions, an anisotropy mesh needs to
be generated during this procedure. Deferent monitor functions for
$x$ and  $y$ coordinates takes
\begin{align*}
\omega_x=\sqrt{1+0.0001|\partial_x(\log\rho)|^2},
\end{align*}
and
\begin{align*}
\omega_y=\sqrt{1+0.01|\partial_y(\log\rho)|^2+0.01|\partial_y\sqrt{U^2+V^2}|}.
\end{align*}
The computational mesh with $100\times50$ cells at $t=0.25, 0.5,
0.75$ and $1$ are given in Fig.\ref{viscous-shock-1} to present the
adaptation procedure. The density distributions on the mesh with
$500\times250$ and $700\times350$  cells are presented in
Fig.\ref{viscous-shock-2} as well. As shown in Tab.\ref{height},
the height of primary vortex predicted by the current scheme agrees
well with the reference data \cite{Case-Kim}. However, due to the
nonlinear combination of linear polynomial, the current scheme is
more dissipative than the fifth-order WENO reconstruction on
structured meshes. Currently, the development of higher-order WENO
on unstructured meshes is in progress.

\begin{figure}[!h]
\centering
\includegraphics[width=0.475\textwidth]{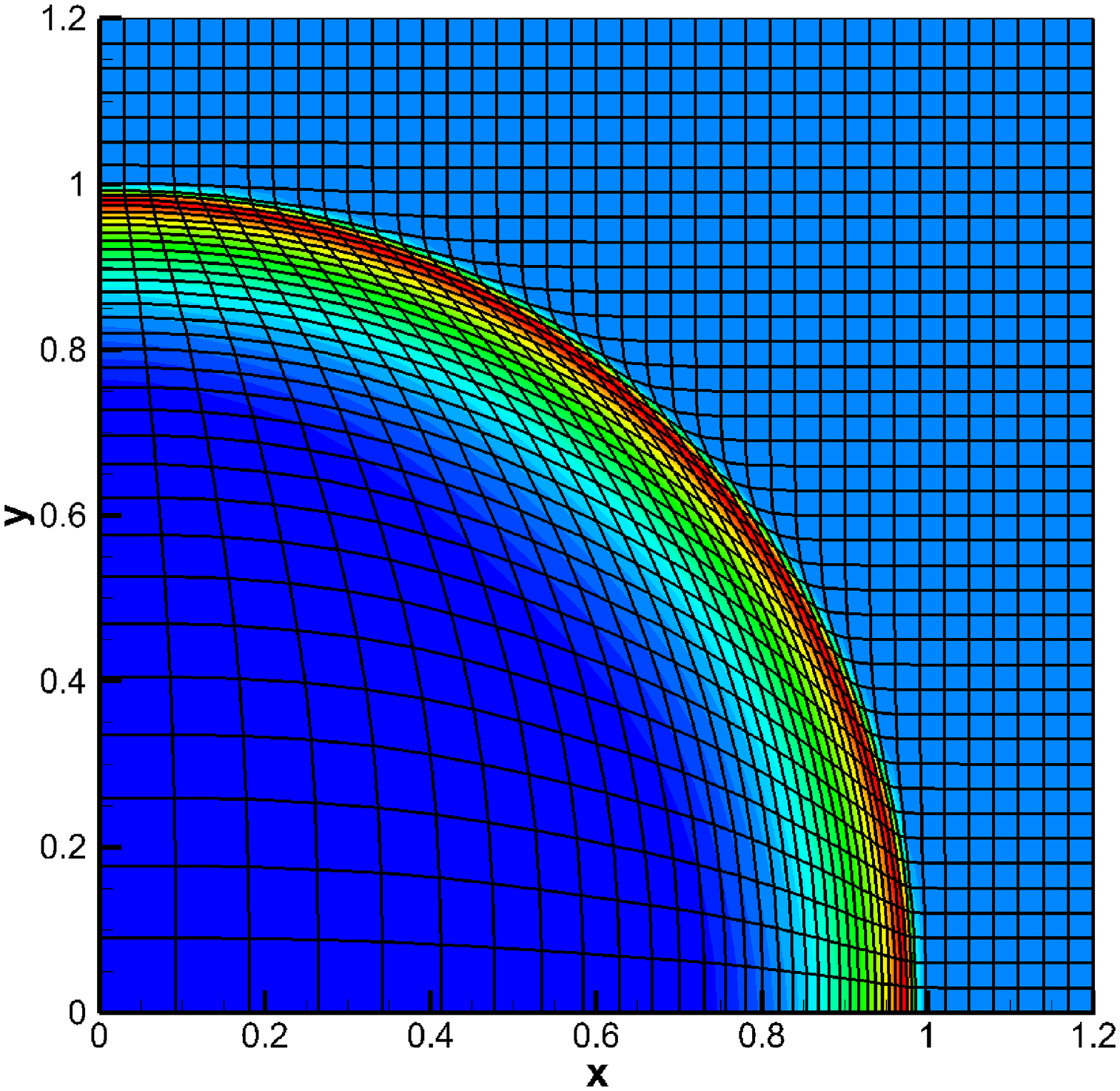}
\caption{\label{Sedov-a} Sedov problem: the mesh and density
distribution at $t=1$ with $40\times40$ cells.}
\includegraphics[width=0.475\textwidth]{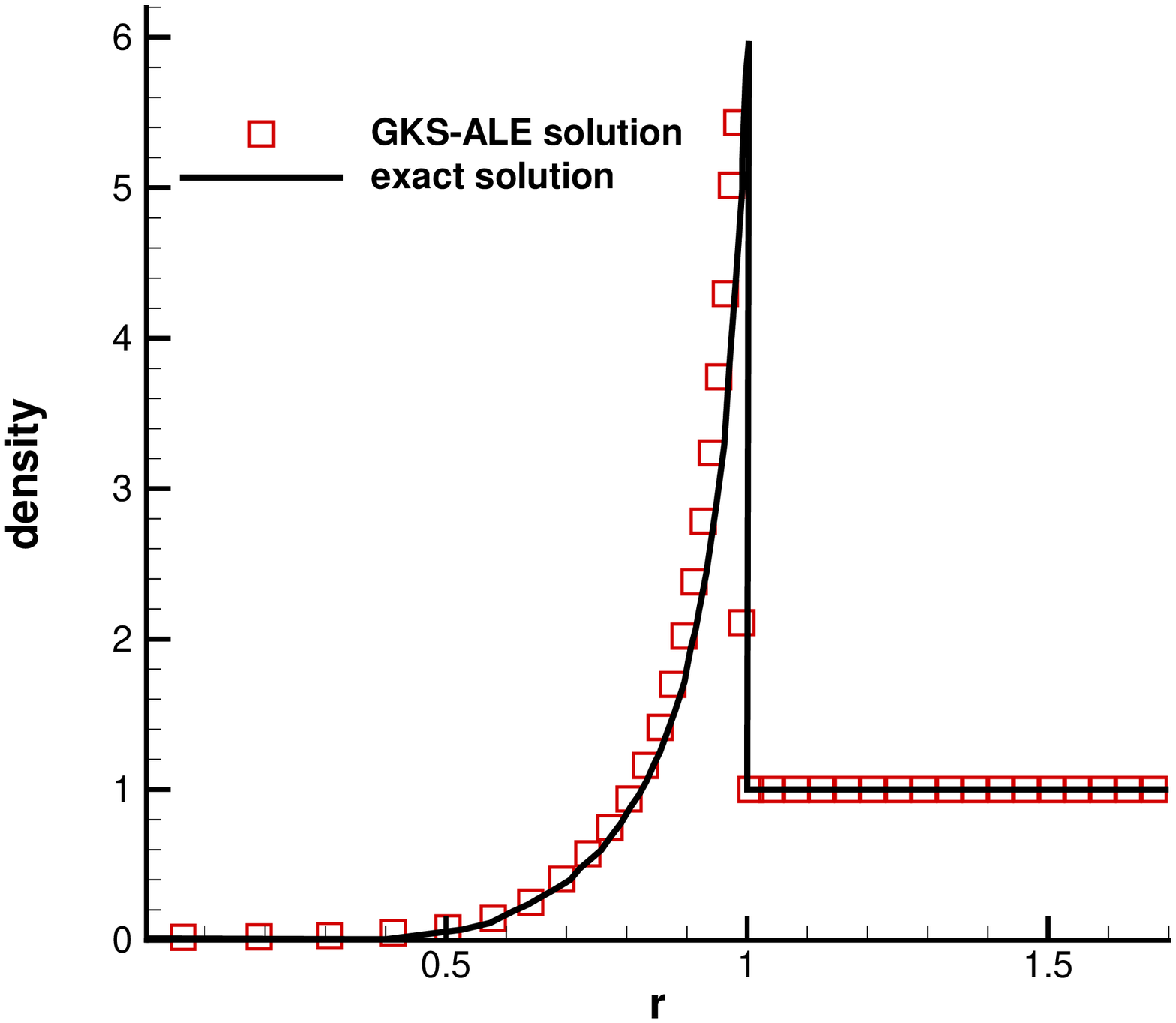}
\includegraphics[width=0.475\textwidth]{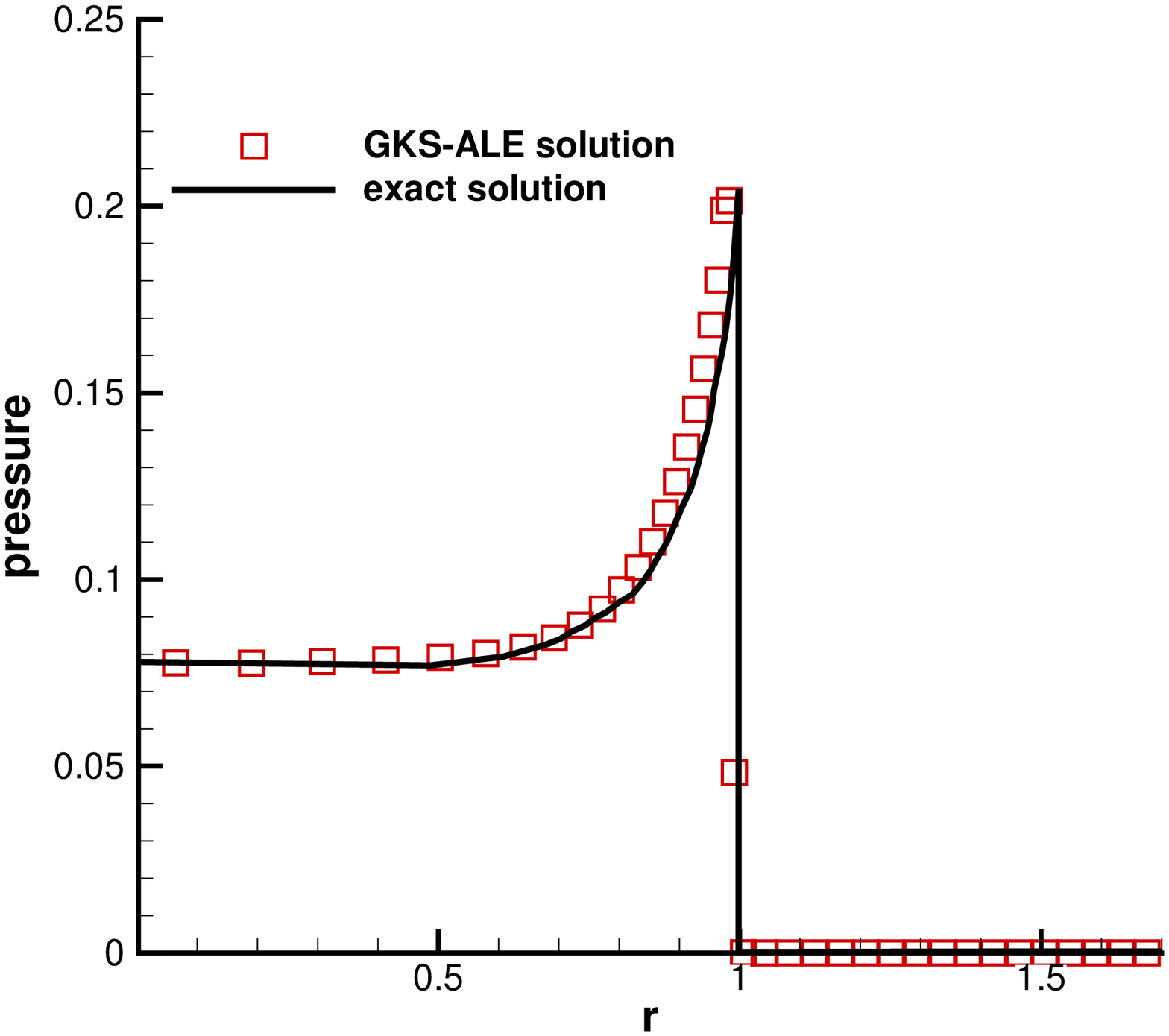}
\caption{\label{Sedov-b} Sedov problem: the density and pressure
distribution along the diagonal line $t=1$.}
\end{figure}

\subsection{Sedov blast wave problem}
This is a two-dimensional explosion problem to model blast wave from
an energy deposited singular point, which is a standard benchmark
problem for the Lagrangian method. The fluid is modeled by the ideal
gas EOS with $\gamma=1.4$. The initial density has a uniform unit
distribution, and the pressure is $10^{-6}$ everywhere, except in
the cell containing the origin. For this cell containing the origin,
the pressure is defined as $p=(\gamma-1)\varepsilon_0/V$, where
$\varepsilon_0=0.244816$ is the total amount of released energy and
$V$ is the cell volume.  The reflection condition is used for the
left and bottom boundaries, and non-reflection condition is used for
the right and top boundaries. The computation domain is
$[0,1.2]\times[0,1.2]$ and $40\times 40$ cells are equally
distributed initially. The density distribution at $t=1$  is given
in Fig.\ref{Sedov-a}. The solution consists of a diverging infinite
strength shock wave whose front is located at radius $r=1$ at $t=1$
\cite{Case-Sedov}.  The density and pressure profile along the
diagonal line $t=1$  are given in Fig.\ref{Sedov-b}, and the
numerical results agrees well with the exact solutions.

\subsection{Saltzman problem}
\begin{figure}[!h]
\centering
\includegraphics[width=0.75\textwidth]{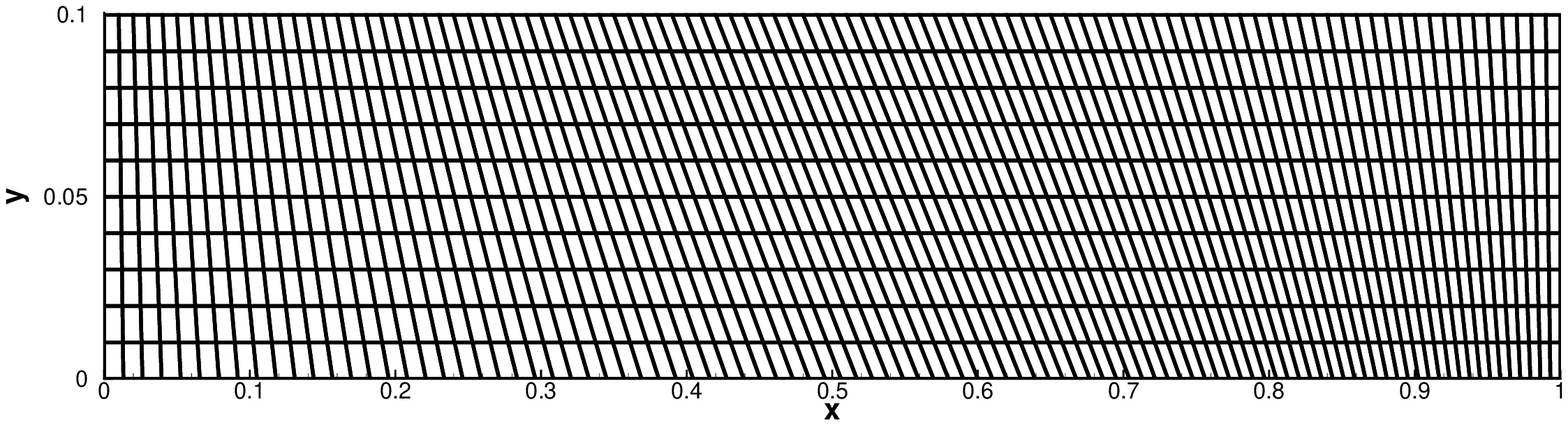}
\includegraphics[width=0.75\textwidth]{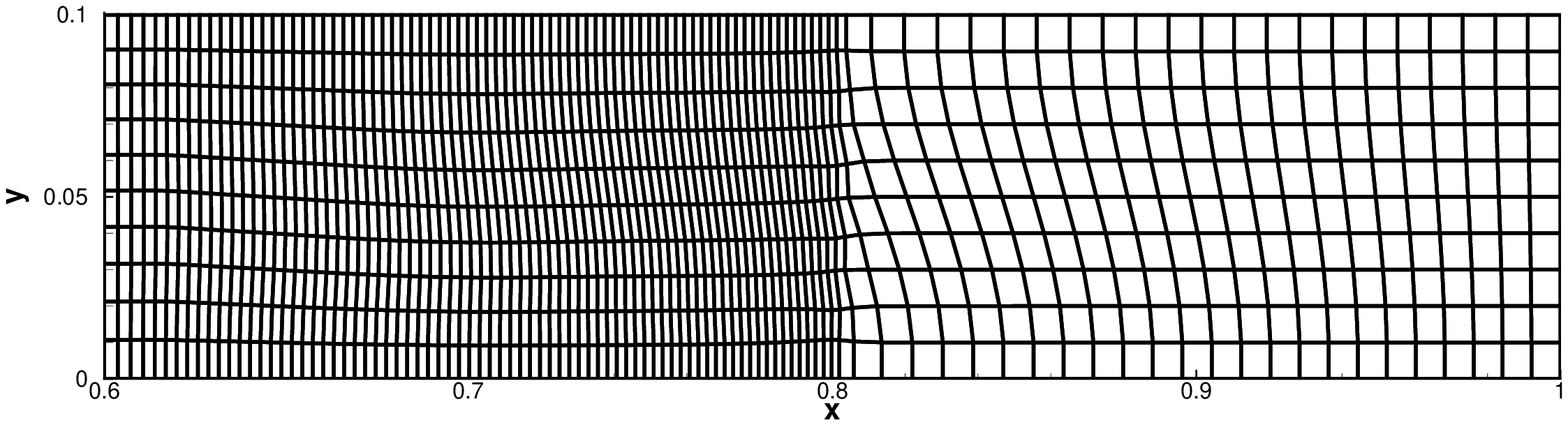}
\caption{\label{Saltzman-a} Saltzman problem: the mesh distribution
at $t=0$ and $0.6$.}
\includegraphics[width=0.49\textwidth]{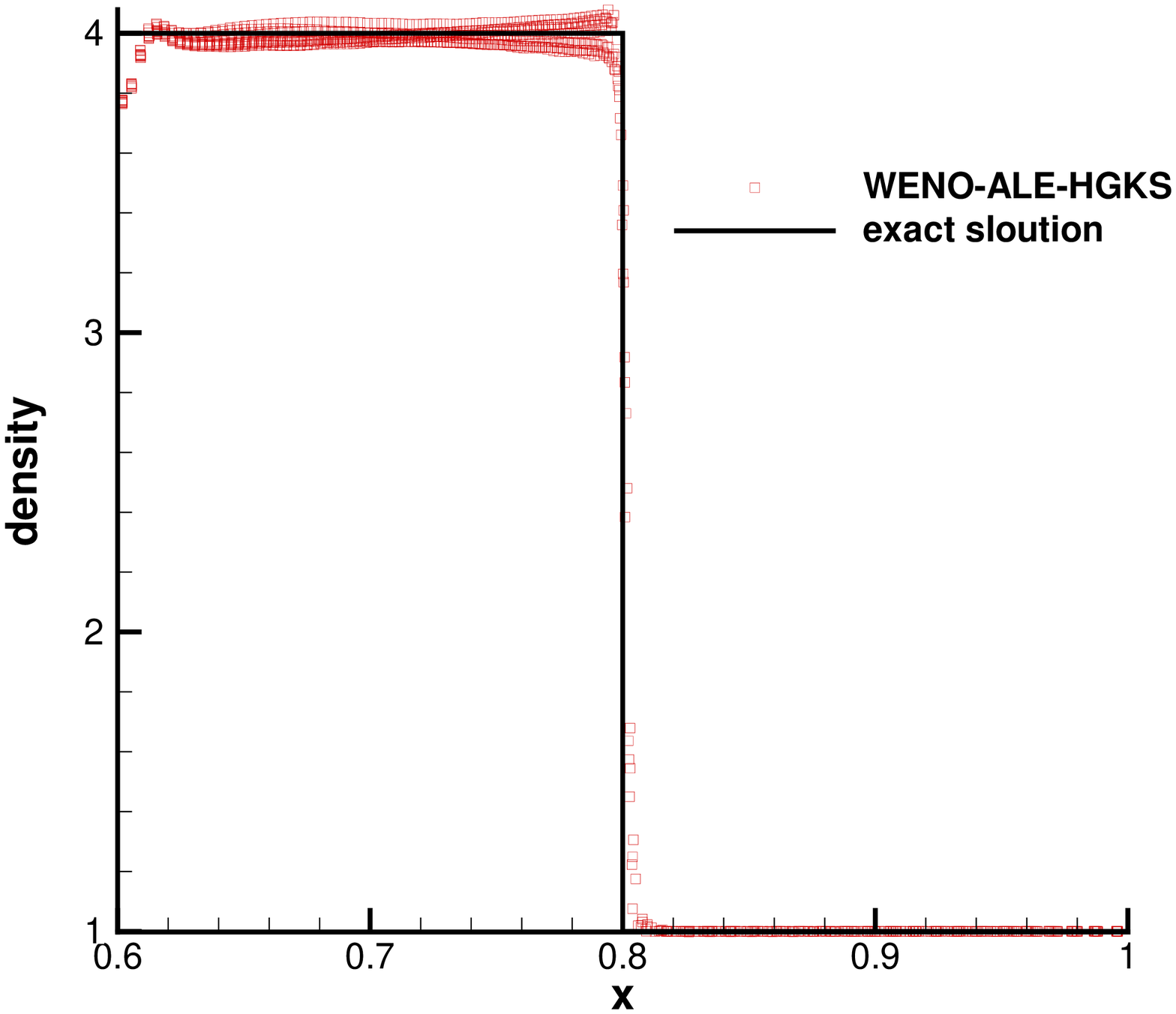}
\includegraphics[width=0.49\textwidth]{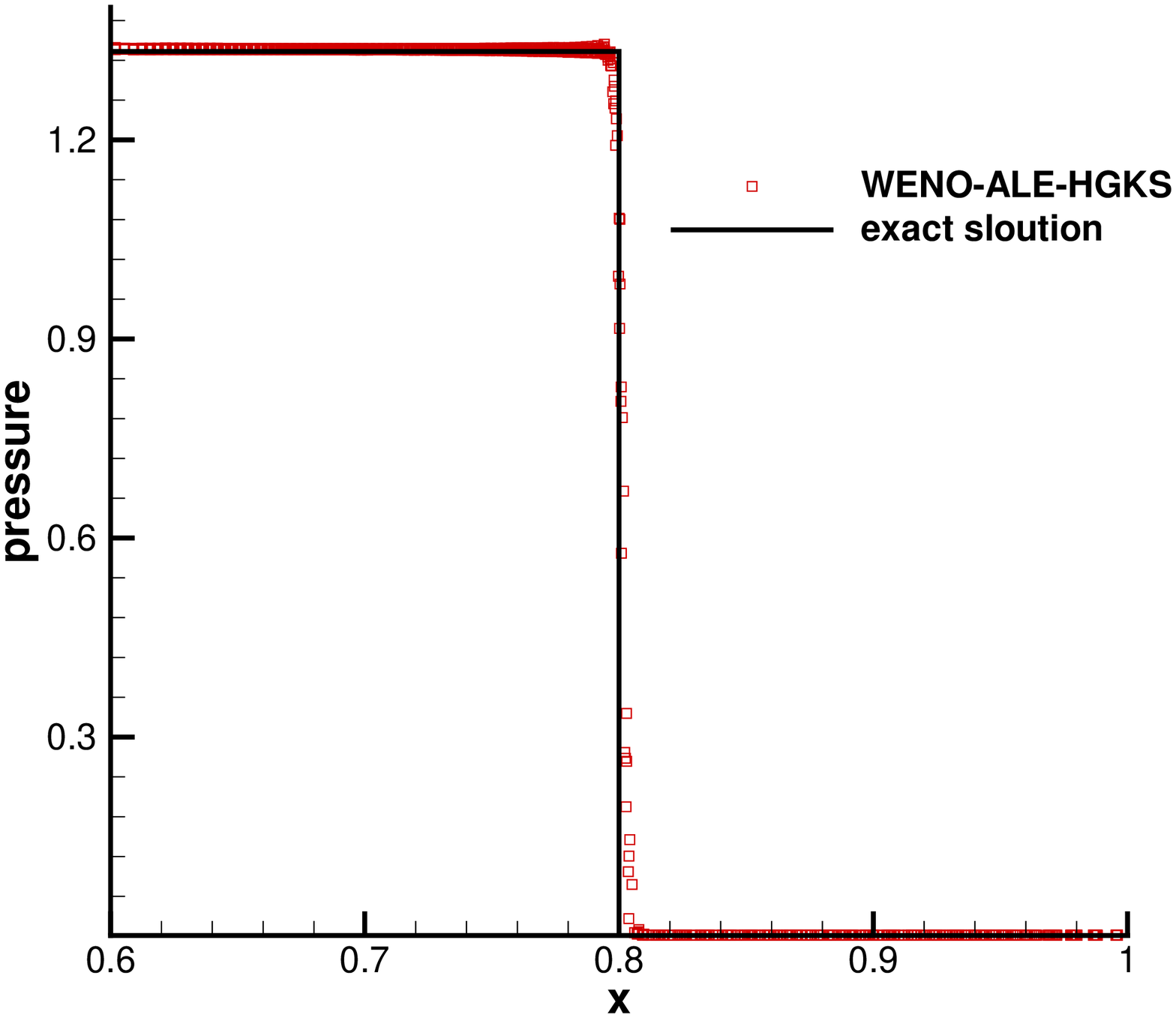}
\caption{\label{Saltzman-b} Saltzman problem: the density and
pressure distribution in all the cells at $t=0.6$.}
\end{figure}

This is another benchmark test case for Lagrangian and ALE codes,
which tests the ability to capture the shock propagation with a
systematically distorted mesh. The computational domain is $[0,
1]\times[0, 0.1]$. The initial mesh is given by
\begin{align*}
x_{ij}&=i\Delta x + (10-j)\Delta y \sin (\frac{i\pi}{100}),\\
y_{ij}&=j\Delta y,
\end{align*}
where $\Delta x=\Delta y=0.01$, and $i=0,...,100, j=0,...,10$. An
ideal monatomic gas with $\rho=1, e=10^{-4}, \gamma=5/3$ is filled
in the box. The left-hand side wall acts as a piston with a constant
velocity $U_p=1$, and other boundaries are reflective walls. As a
consequence, a strong shock wave is generated from the left end. At
time $t=0.6$, the shock is expected to be located at $x=0.8$, and
the post shock solutions are $p=4$ and $p=1.333$. The meshes
distribution at $t=0$ and $0.6$ are given in Fig.\ref{Saltzman-a}.
The density and pressure distribution in all the cells at $t=0.6$ is
given in Fig.\ref{Saltzman-b}. The numerical results agrees well
with the exact solutions.

\section{Conclusion}
In this paper, a gas-kinetic scheme is developed based on the
third-order WENO reconstruction on unstructured quadrilateral
meshes, in which the optimization approach for linear weights and
the non-linear weights with new smooth indicator are proposed to
improve the robustness of reconstruction. With the arbitrary
Lagrangian-Eulerian (ALE) formulation, a high-order moving-mesh
gas-kinetic scheme is presented for the inviscid and viscous flows.
It extends the third-order gas kinetic method from a static domain
to the flow simulation over a variable domain. The two-stage
fourth-order method is used for the temporal discretization, and the
second-order gas-kinetic solver is used for the flux calculation. In
the two-stage method, the spatial reconstruction is performed at the
initial and intermediate stage, while the mesh at corresponding
stage is given through a specified mesh velocity. To take the mesh
velocity along each cell interface into account, the WENO
reconstruction is performed at each Gaussian quadrature point in the local
moving coordinate. Therefore, the geometric conservation law can be
satisfied accurately, which is also verified numerically even with
the largely deforming mesh. Numerical examples are presented to
validate the performance of current scheme, where the mesh
adaptation method and cell centered Lagrangian method can be used to
provide mesh velocity.

\section*{Acknowledgements}
The current research of L. Pan is supported by National Science
Foundation of China (11701038), the Fundamental Research Funds for
the Central Universities and a grant from the Science $\&$
Technology on Reliability $\&$ Environmental Engineering Laboratory
(No. 6142A0501020317). The work of K. Xu is supported by National
Science Foundation of China (11772281, 91852114) and Hong Kong
research grant council (16206617).

\end{document}